\documentclass[11pt,aps,a4paper,eqsecnum,amsmath,amssymb
              ,nofootinbib
              ,superscriptaddress,notitlepage%,endfloats*
              ]{revtex4}

\usepackage{graphicx}
\newcommand{\exponent}{\delta}	% Exponent in L^exponent
\newcommand{\Lambdamix}{\Lambda_{\textrm{mix}}}
\newcommand{\Omegamix}{\Omega_{\textrm{mix}}}
\newcommand{\Lambdamixstag}{\Lambda^{(2)}}
\newcommand{\kah}{t}
\renewcommand{\ln}{\log}		% alle lns sind logs
\newcommand{\I}{\mathrm{i}}        % imagin"are Einheit
\newcommand{\E}{\mathrm{e}}        % e Eulersche Zahl
\newcommand{\D}{\mathrm{d}}        % Differential

\newcommand{\End}{\operatorname{End}}   %Endomorphismus
\renewcommand{\Im}{\operatorname{Im}} % Imaginaerteil
\newcommand{\ch}{\operatorname{ch}}  % Cosinus-hyperbolicus
\newcommand{\sh}{\operatorname{sh}}

\newcommand{\half}{\frac{1}{2}}

\begin{document}

\preprint{}
\title{The Staggered Six-Vertex Model: Conformal Invariance\newline
  and Corrections to Scaling}

\author{Holger Frahm}
\affiliation{%
Institut f\"ur Theoretische Physik, Leibniz Universit\"at Hannover,
Appelstra\ss{}e 2, 30167 Hannover, Germany}

\author{Alexander Seel}
\affiliation{%
Lehrstuhl f\"ur Theoretische Elektrotechnik und Photonik, Universit\"at Siegen, 
H\"olderlinstra\ss{}e 3, 57068 Siegen, Germany}

%\date{\today}

\begin{abstract}
  We study the emergence of non-compact degrees of freedom in the low energy
  effective theory for a class of $\mathbb{Z}_2$-staggered six-vertex models.
  In the finite size spectrum of the vertex model this shows up through the
  appearance of a continuum of critical exponents.  To analyze this part of
  the spectrum we derive a set of coupled nonlinear integral equations from
  the Bethe ansatz solution of the vertex model which allow to compute the
  energies of the system for a range of anisotropies and of the staggering
  parameter.  The critical theory is found to be independent of the
  staggering.  Its spectrum and density of states coincide with the
  $SL(2,\mathbb{R})/U(1)$ Euclidean black hole conformal field theory which
  has been identified previously in the continuum limit of the vertex model
  for a particular 'self-dual' choice of the staggering.  We also study the
  asymptotic behaviour of subleading corrections to the finite size scaling
  and discuss our findings in the context of the conformal field theory.
\end{abstract}

\maketitle

%%%%%%%%%%%%%%%%%%%%%%%%%%%%%%%%%%%%%%%%%%%%%%%%%%%%%%%%%%%%%%%%%%%%%% 
\section{Introduction}
%%%%%%%%%%%%%%%%%%%%%%%%%%%%%%%%%%%%%%%%%%%%%%%%%%%%%%%%%%%%%%%%%%%%%%
%
Studies of two-dimensional vertex models with local Boltzmann weights
satisfying the Yang-Baxter equation or the related quantum spin chains have
provided tremendous insights into the properties of quantum field theories in
$(1+1)$ dimension.  The solution of the lattice model by means of Bethe ansatz
methods allows a complete characterization of the spectrum in terms of
elementary excitations and the analysis of the finite size spectrum allows for
the identification of the model as a lattice regularization of the low
energy effective field theory.

Based on a particular solution of the Yang-Baxter equation vertex models on
lattices of arbitrary sizes can be defined using the co-multiplication
property of the Yang-Baxter algebra.  Introducing inhomogeneities -- either by
shifts in the spectral parameter and/or by using different representations of
the underlying algebra for the internal local degrees of freedom
\cite{Baxt71,KuRS81} -- allows to generalize these models further while
keeping their integrability or to uncover relations to different systems: the
equivalence of the $q$-state Potts model on the square lattice and a staggered
six-vertex model \cite{TeLi71,BaKW76} allowed for the solution of the Potts
model by means of the Bethe ansatz \cite{Baxt82}.  Other choices for the
staggering parameter in the six-vertex model lead to integrable quantum chains
with longer ranged interactions \cite{PoZv93,FrRo96} and have been found to
appear in the zero charge sector of a vertex model based on alternating
four-dimensional representations of the quantum group deformation of the Lie
superalgebra $sl(2|1)$ \cite{FrMa12}.

Recently, the critical properties of the staggered six-vertex model have been
investigated for a particular choice of the staggering parameter corresponding
to one of the integrable manifolds of the antiferromagnetic $q$-state Potts
model \cite{JaSa06,IkJS08}.  For this 'self-dual' case the model has an
additional discrete $\mathbb{Z}_2$-invariance leading to a conserved charge
which can be used to classify the spectrum.  Remarkably, it has been found
that although the lattice model is defined in terms of (compact) spin-$1/2$
degrees of freedom the low energy effective theory has a continuous spectrum
of critical exponents.  From a finite size scaling analysis and the
computation of the density of states in the continuum the latter has been
identified with the non-compact $SL(2,\mathbb{R})_k/U(1)$ sigma model, a
conformal field theory (CFT) on the two-dimensional Euclidean black hole
background \cite{IkJS12,CaIk13}.  The appearence of a non-compact continuum
limit of a lattice model with finite number of states per site has also been
observed in staggered vertex models with supergroup symmetries, see e.g.\
Refs.~\onlinecite{EsFS05,Cand11,FrMa11,FrMa12}.

The spectrum for finite lattices is necessarily discrete.
In the staggered six-vertex model the continuous spectrum emerges in the
thermodynamic limit by closing the gaps between critical exponents as
$1/(\ln L)^2$.  As a consequence the investigation of this scenario requires
to compute energies for very large system sizes.  For the integrable models
mentioned above this is only possible after formulation of the spectral
problem in terms of nonlinear integral equations (NLIEs) in which the system
size enters as a parameter only.  For the self-dual staggered six-vertex model
such NLIEs have been derived and solved numerically to identify the low energy
effective theory \cite{CaIk13}.

In this paper we derive a different set of NLIEs for the finite size spectrum
of the spin chain from the Bethe ansatz solution of the staggered six-vertex
model with anisotropy $0<\gamma<\pi/2$.  These equations hold for arbitrary
values of the staggering parameter $\gamma<\alpha<\pi-\gamma$, in particular
away from the self-dual line $\alpha=\pi/2$.  After recalling what is known
about the finite size spectrum of this model we propose a parametrization of
its emerging continuous part in terms of the eigenvalues of the quasi-momentum
operator.  We solve the NLIEs numerically for systems with up to $10^6$
lattice sites to verify this proposal.  Both the finite size spectrum and the
density of states are found to depend only on the anisotropy $\gamma$ but not
on the staggering parameter $\alpha$.  This suggests that the effective theory
for the staggered model is the black hole sigma model CFT
$SL(2,\mathbb{R})_k/U(1)$ at level $k=\pi/\gamma>2$, independent of $\alpha$.
We also study the subleading terms appearing in the finite size scaling of the
energies of the system and propose a possible explanation for our findings in
the context of this CFT.

\section{The staggered six-vertex model}
\begin{figure}[t]
\begin{center}
\includegraphics[height=7cm]{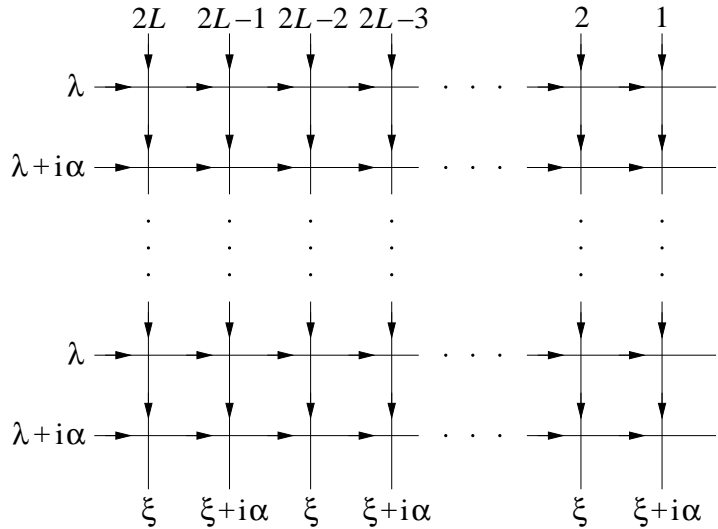}
\end{center}
\caption{The staggered six-vertex model on a square lattice. The local
  Boltzmann weights are elements of the $R$-matrix related to the six-vertex
  model.}
\label{vertices}
\end{figure}

The six-vertex model on the square lattice, see Fig.~\ref{vertices},
%
%======================================================================%
%======================================================================%
%
is defined through the $R$-matrix
\begin{equation}
\label{R6v}
R_{12}(\lambda,\mu) = \begin{pmatrix}
                    a(\lambda,\mu) & 0 & 0 & 0 \\
		    0 & b(\lambda,\mu) & c(\lambda,\mu) & 0 \\[.5ex]
		    0 & c(\lambda,\mu) & b(\lambda,\mu) & 0 \\
		    0 & 0 & 0 & a(\lambda,\mu)
 \end{pmatrix} \quad , \quad 
\end{equation}
containing the local Boltzmann weights $a(\lambda,\mu)\equiv 1$,
$b(\lambda,\mu) = \frac{\sh(\lambda-\mu)}{\sh(\lambda-\mu+\I\gamma)} $,
$c(\lambda,\mu) = \frac{\sh(\I\gamma)}{\sh(\lambda-\mu + \I\gamma)}$.  
The matrix $R_{12}(\lambda,\mu)$ should be read as an element of $\End
V_1\otimes\End V_2$, $V_i\simeq\mathbb{C}^2$.  The spin variables lying on the
horizontal and vertical links of the lattice take values from $V_1$ and $V_2$,
respectively. Similarly, we associate the spectral parameters $\lambda$ and
$\mu$ to horizontal and vertical lines while $\gamma$ parametrizes the
anisotropy of the model.

Taking the trace of an ordered product of $R$-matrices we construct a family
of row-to-row transfer matrices on a lattice with $2L$ horizontal sites with
staggered spectral parameters
\begin{equation}
\label{trans1}
  t(\lambda) = \mathrm{tr}_0\left( R_{0,2L}(\lambda,\xi)
    R_{0,2L-1}(\lambda,\xi+\I\alpha) \ldots 
    R_{0,2}(\lambda,\xi)R_{0,1}(\lambda,\xi+\I\alpha)\right)\,. 
\end{equation}
The fact that the $R$-matrix (\ref{R6v}) satisfies the Yang-Baxter equation
implies that these transfer matrices commute for arbitrary values of
$\lambda$, i.e.\ $[t(\lambda),t(\mu)]=0$.

In the vertex model depicted in Fig.~\ref{vertices} we have introduced an
additional staggering in the vertical direction.  The corresponding double-row
transfer matrix is the product of commuting operators
\begin{equation}
\label{trans2}
  t^{(2)}(\lambda) = t(\lambda)\,t(\lambda+\I\alpha)\;.
\end{equation}
On two lines in the space of coupling constants $\gamma$ and $\alpha$ this
staggered six-vertex model exhibits additional quantum group symmetries: 
the first occurs for $\alpha=\gamma$.  In this case the product of
neighbouring $R$-matrices in (\ref{trans1}) degenerates and the underlying
quantum group symmetry is that present in the integrable spin-1 XXZ chain
\cite{ZaFa80}.
A second special line follows from the spectral equivalence of the models with
staggering parameter $\alpha$ and $\pi-\alpha$ \cite{FrMa12}.  This leads to
the presence of a discrete $\mathbb{Z}_2$-invariance on the 'self-dual' line
$\alpha=\pi/2$ where the model is equivalent to one of the integrable
manifolds of the antiferromagnetic $q$-state Potts model \cite{Baxt82,IkJS08}.
On this line the model has a quantum group symmetry related to the twisted
quantum algebra $U_q[D_2^{(2)}]$ with $q=\E^{2\I\gamma}$ \cite{FrMa12}.

%%%%%%%%%%%%%%%%%%%%%%%%%%%%%%%%%%%%%%%%%%%%%%%%%%%%%%%%%%%%%%%%%%%%%%
The commuting operators generated by the double-row transfer matrix
(\ref{trans2}) can be written as sums over local (i.e.\ finite-range on the
lattice) interactions, independent of the system size: as a consequence of
$R(\lambda,\lambda)$ 
%$R(\lambda,\lambda)\equiv\mathcal{P}$ 
being the permutation operator on $V_1\otimes V_2$,
$t^{(2)}(\xi)$ acts as a two-site translation operator. Therefore we can
define the momentum operator as $P=-\I\ln
t^{(2)}(\xi)=-\I\ln(t(\xi)\,t(\xi+\I\alpha))$.  The next term in the expansion
of $\ln t^{(2)}(\xi)$ is the spin chain Hamiltonian
\begin{equation}
  \label{hamiltonian}
  H \equiv -\I\partial_\lambda
  \left. \ln\left(t(\lambda)\,t(\lambda+\I\alpha)\right) \right|_{\lambda=\xi}
\end{equation}
with nearest and next-nearest neighbour interaction. In terms of local Pauli
matrices the Hamiltonian reads up to an overall factor
\begin{equation}
\label{hamiltonian3}
\begin{aligned} 
%\frac{\sin(\alpha+\gamma)\sin(\alpha-\gamma)\sin\gamma}{\cos\gamma}\cdot 
  H \propto&\sum_{j=1}^{2L}\bigg[ -\frac{\sin^2\!\alpha}{2}\,
    \big(\boldsymbol\sigma_j\cdot\boldsymbol\sigma_{j+2}\big)
  +%\sum_{j=1}^{2L}&
  \sin^2\!\gamma \bigg\{ 
  \frac{\cos\alpha}{\cos\gamma}
  \big(\sigma_j^x\sigma_{j+1}^x+\sigma_j^y\sigma_{j+1}^y\big)+
  \sigma_j^z\sigma_{j+1}^z\bigg\} \\
  &\qquad+\I(-1)^j\frac{\sin\gamma\sin\alpha}2
  \big(\sigma^x_{j}\sigma^y_{j+1}-\sigma^y_{j}\sigma^x_{j+1}\big)
  \big(\sigma_{j-1}^z-\sigma_{j+2}^z\big)\\[1ex]
  &\qquad+\I(-1)^j\frac{\sin\gamma\sin(2\alpha)}{4\cos\gamma}\,
  \big(\sigma^x_{j-1}\sigma^y_{j+1}-\sigma^y_{j-1}\sigma^x_{j+1}\big)\,\sigma_{j}^z
  \bigg]\\
    &+L\Big(\cos(2\gamma)-\cos^2\!\alpha\Big)
\; .
\end{aligned}
\end{equation}
Note that this Hamiltonian differs from the one given by Ikhlef \emph{et al.}
\cite{IkJS08} for $\alpha=\pi/2$ by a unitary rotation of spins on one
sublattice.
Following Refs.~\onlinecite{IkJS12,CaIk13} we also define a \emph{quasi-shift}
operator $\widetilde{\tau} \equiv t(\xi)\left[t(\xi+\I\alpha)\right]^{-1}$ and
the corresponding \emph{quasi-momentum}
\begin{equation}
  \label{quasimomentum}
  \widetilde{P} \equiv \ln\left({t(\xi)}\,[t(\xi+\I\alpha)]^{-1} \right)\,.
\end{equation}
Similar as in the self-dual case $\widetilde{\tau}$ acts as a
diagonal-to-diagonal (light-cone) transfer matrix.

%%%%%%%%%%%%%%%%%%%%%%%%%%%%%%%%%%%%%%%%%%%%%%%%%%%%%%%%%%%%%%%%%%%%%%
\section{Bethe ansatz solution}
\label{sec:BAthermo}
%%%%%%%%%%%%%%%%%%%%%%%%%%%%%%%%%%%%%%%%%%%%%%%%%%%%%%%%%%%%%%%%%%%%%%
In the six-vertex model the number of down arrows is conserved.  This $U(1)$
symmetry allows to diagonalize the transfer matrix starting from the reference
state $|0\rangle\equiv\big(\begin{smallmatrix}1\\
  0\end{smallmatrix}\big)^{\otimes 2L}$ by means of the algebraic Bethe ansatz
\cite{VladB}.  The resulting eigenvalues in the sector with $M\le L$
down arrows are
\begin{equation}
\label{basiceigenvalue}
  \Lambda(\lambda)=
  \prod_{\ell=1}^M \frac{\sh(\lambda-\lambda_\ell-\I\gamma)}{
    \sh(\lambda-\lambda_\ell)} + 
  d(\lambda)\, \prod_{\ell=1}^M \frac{\sh(\lambda-\lambda_\ell+\I\gamma)}{
    \sh(\lambda-\lambda_\ell)}\,,
\end{equation}
%model $a(\lambda)\equiv1$
with
\begin{equation}
  d( \lambda)\equiv\left(\frac{\sh(\lambda-\xi)}{\sh(\lambda-\xi+\I\gamma)}
    \frac{\sh(\lambda-\xi-\I\alpha)}{\sh(\lambda-\xi-\I\alpha+\I\gamma)}
  \right)^L  \,. 
\end{equation}
In Eq.~(\ref{basiceigenvalue}) the rapidities $\lambda_j$ are solutions to the
Bethe equations
\begin{equation} 
  d(\lambda_j)\,
  \prod_{\substack{\ell=1\\ \ell\not=j}}^M
  \frac{\sh(\lambda_j-\lambda_\ell+\I\gamma)}{
    \sh(\lambda_j-\lambda_\ell-\I\gamma)} = 1\,,
   \quad j=1,\ldots,M  \, .
\end{equation}
Note that these equations ensure the analyticity of the transfer matrix
eigenvalues $\Lambda(\lambda)$ at the points $\lambda=\lambda_j$,
$j=1,\ldots,M$. 

Since the expressions above depend only on the difference $\lambda-\xi$ we are
free  to choose $\xi\equiv\I(\gamma-\alpha)/2$.  This results in the symmetric
Bethe equations
\begin{equation} 
\label{BAequ}
  \left(\frac{\sh(\lambda_j+\frac{\I\alpha}2+\frac{\I\gamma}2)}
    {\sh(\lambda_j+\frac{\I\alpha}2-\frac{\I\gamma}2)}
    \frac{\sh(\lambda_j-\frac{\I\alpha}2+\frac{\I\gamma}2)}
    {\sh(\lambda_j-\frac{\I\alpha}2-\frac{\I\gamma}2)}\right)^L =
  \prod_{\substack{\ell=1\\ \ell\not=j}}^M
  \frac{\sh(\lambda_j-\lambda_\ell+\I\gamma)}{
    \sh(\lambda_j-\lambda_\ell-\I\gamma)} \,, \quad j=1,\ldots,M \;.
\end{equation}

The eigenvalues of the double-row transfer matrix (\ref{trans2}) are the
product of the individual ones from Eq.~(\ref{basiceigenvalue}):
\begin{equation}
 \Lambdamixstag(\lambda) = \Lambda(\lambda)\,\Lambda(\lambda+\I\alpha)\,.
\end{equation}
Similarly, the energy eigenvalues $E$ of the Hamiltonian \eqref{hamiltonian}
and the eigenvalue $K$ of the quasi-momentum operator \eqref{quasimomentum}
are found to be:
\begin{equation} 
\label{finaleqs}
  E=-\I\partial_\lambda \left.\log\left(\Lambda(\lambda)\,
  \Lambda(\lambda+\I\alpha)\right)\right|_{\lambda=\xi}\,, \quad
  K= \log\left(\Lambda(\xi)/\Lambda(\xi+\I\alpha)\right)\,.
\end{equation}
Using the expression (\ref{basiceigenvalue}) for $\Lambda(\lambda)$ in terms
of the Bethe roots $\{\lambda_j\}$ they are found to be sums over
contributions from single rapidities, i.e.\ $E=\sum_{j=1}^{M}
\epsilon_0(\lambda_j)$ and $K = \sum_{j=1}^{M} k_0(\lambda_j)$, where
\begin{equation}
\begin{aligned}
%  E&=\sum_{\ell=1}^{L} \epsilon(\lambda_\ell)\,,\quad 
  \epsilon_0(\lambda) &=
  \frac{4\sin\gamma\big[\ch(2\lambda)\cos\alpha-\cos\gamma\big]} 
       {\big[\ch(2\lambda)-\cos(\alpha-\gamma)\big]
         \big[\ch(2\lambda)-\cos(\alpha+\gamma)\big]}  \,,
       \\[1ex]
%       K &=       \sum_{\ell=1}^{L} 
       k_0(\lambda) &=
       \log\bigg(\frac{\ch(2\lambda)-\cos(\alpha+\gamma)}{
         \ch(2\lambda)-\cos(\alpha-\gamma)} \bigg) \,.
\end{aligned}
\end{equation}

In the following we shall analyze the ground state and lowest excitations of
the model with Hamiltonian (\ref{hamiltonian}) for staggering
$\gamma<\alpha<\pi-\gamma$ (denoted as phase B in Ref.~\onlinecite{FrMa12}).
In this regime the low energy states have been found to be described by
configurations involving two types of Bethe roots for all anisotropies
$0\le\gamma\le\pi/2$, namely $\{\lambda_j\}$ with $\mathrm{Im}(\lambda_j)=0$
or $\pi/2$:
\begin{gather} 
\label{Bethe-roots} 
\{\lambda_\ell\}_{\ell=1}^M =
  \{\nu_j\}_{j=1}^{n_1} \cup \{\frac{\I\pi}{2}+\mu_j\}_{j=1}^{n_2}
  \quad\text{with}\quad
  \nu_j,\mu_j\in\mathbb{R}\,.
\end{gather}
Among these the ground state of the staggered spin chain is realized in the
sector with $M=n_1+n_2=L$ and roots $\nu_j$, $\mu_j$ filling the entire real
axis.  Taking the thermodynamic limit $L\to\infty$ with fixed $n_1/L$, $n_2/L$
their distributions are described by densities $\rho_1(\nu)$ and $\rho_2(\mu)$
which are determined through coupled linear integral equations
\cite{YaYa69,HUBBARD}. In the range of parameters considered here these
equations have been solved to give \cite{FrMa12}
\begin{equation}
  \label{densB}
  \begin{aligned}
    \rho_1(\nu) &= 
       \frac{\sin\frac{\pi(\alpha-\gamma)}{\pi-2\gamma}}{2(\pi-2\gamma)}\,
       \left(\cosh\frac{2\pi \nu}{\pi-2\gamma} -
         \cos\frac{\pi(\alpha-\gamma)}{\pi-2\gamma}\right)^{-1}\,,\\[1ex] 
    \rho_2(\mu) &=
       \frac{\sin\frac{\pi(\alpha-\gamma)}{\pi-2\gamma}}{2(\pi-2\gamma)}\,
       \left(\cosh\frac{2\pi \mu}{\pi-2\gamma} +
         \cos\frac{\pi(\alpha-\gamma)}{\pi-2\gamma}\right)^{-1}\,.
  \end{aligned}
\end{equation}
Integrating these expressions we find the total densities of the two types of
roots in the ground state to be
\begin{equation}
  \label{TLstaggering}
  \frac{n_1^{(0)}}{L} = \int_{-\infty}^{\infty} \D\nu\, \rho_1(\nu) =
  \frac{\pi-\alpha-\gamma}{\pi-2\gamma}     =1-\frac{n_2^{(0)}}{L}\, .
\end{equation}
Note that under the action of the duality transform $\alpha\to\pi-\alpha$ the
two types of roots are exchanged.  This allows to restrict our analysis to
staggering parameters $\gamma<\alpha\le\pi/2$ in the following.

The low energy excitations of the model have a linear dispersion and can be
characterized by the deviations of the numbers $n_{1,2}$ from their ground
state values (\ref{TLstaggering}), i.e.\
\begin{equation}
\label{BAnos}
  \Delta n_{1,2} = n_{1,2}-n_{1,2}^{(0)} \equiv \half\left(m \pm
    \widetilde{m}\right) \,
\end{equation}
(note that $m=n_1+n_2-L\in\mathbb{Z}$ by construction), and their momentum
which can be parametrized by a single vorticity $w\in\mathbb{Z}$.  Also within
the root density approach the energy of these excitations has been found to be
\cite{FrMa12} (see also Refs.~\onlinecite{JaSa06,IkJS08} for the self-dual
case)
\begin{equation}
  \label{fse}
  E(L) = L\varepsilon_\infty + \frac{2\pi\,v_F}{L} \left(
    -\frac{1}{6} +  
    \frac{\gamma}{2\pi}\, m^2 + \frac{\pi}{2\gamma}\, w^2 + \kappa(L)\,
    \widetilde{m}^2 + n_+ + n_-  + R(L) \right)\,.
\end{equation}
Here $\varepsilon_\infty$ is the bulk energy density of the state with root
densities (\ref{densB})
\begin{equation}
  \label{ebulk}
  \varepsilon_\infty = -2 \int\displaylimits_{-\infty}^{\infty}\!\!
  \D k\,\frac{\sh\big(\frac{k\gamma}{2}\big)
    \big[\sh\big(\frac{k\pi}{2}-\frac{k\gamma}{2}\big)
    \ch\big(\frac{k\pi}{2}-k\alpha\big)- 
    \sh\big(\frac{k\gamma}{2}\big)\big]}
  {\sh\big(\frac{k\pi}{2}\big)\sh\big(\frac{k\pi}{2}-k\gamma\big)}\,,
\end{equation}
$v_F = 2\pi/(\pi-2\gamma)$ is the Fermi velocity of the low energy modes,
$n_\pm$ are non-negative integers characterizing particle-hole type
excitations and the corrections to scaling $R(L)$ vanish as $L\to\infty$.

Similarly, the eigenvalue $K$ of the quasi-momentum operator can be computed
within the root density approach.  For the ground state $K$ is proportional to
the system size, the corresponding 'quasi-momentum density' of the ground
state in the thermodynamic limit is
\begin{equation}
  \begin{aligned}
  \label{kbulk}
  k_\infty = \frac{1}{L}\,K &= \int_{-\infty}^{\infty} \D\nu\,\rho_1(\nu)k_0(\nu)
          + \int_{-\infty}^{\infty} \D\mu\,\rho_2(\mu)k_0(\mu+\I\pi/2) \\[1ex]
  &= 2 \int\displaylimits_{-\infty}^{\infty}\!\!\D k\,
  \frac{\sh\big(\frac{k\gamma}{2}\big)
    \sh\big(\frac{k\pi}{2}-k\alpha\big)
    \sh\big(\frac{k\pi}{2}-\frac{k\gamma}{2}\big)} 
  {k\,\sh\big(\frac{k\pi}{2}\big)\sh\big(\frac{k\pi}{2}-k\gamma\big)}\,.
  \end{aligned}
\end{equation}
Note that $k_\infty=0$ on the self-dual line $\alpha=\pi/2$ as a consequence
of $k_0(\lambda) =-k_0(\lambda+\I\pi/2)$ and
$\rho_1(\lambda)\equiv\rho_2(\lambda)$. 

To analyze the finite size spectrum (\ref{fse}) further and to identify the
low energy effective theory for the staggered six-vertex model one has to study
the system size dependence of the coupling constant $\kappa(L)$.  Based on
numerical solutions of the Bethe equations (\ref{BAequ}) for systems with
several thousand lattice sites it has been established that $\kappa(L)\propto
1/(\log L)^2$ for large $L$ \cite{JaSa06,IkJS08} indicating the emergence of a
continuous component in the spectrum of critical exponents of the model in the
thermodynamic limit.
Due to this logarithmic dependence on the system size a quantitative analysis
of the spectrum requires more sophisticated methods.   In a first step  Ikhlef
\emph{et al.}  have refined the root density approach using Wiener-Hopf
methods which allowed them to derive  an analytical expression for the
asymptotic behaviour of $\kappa(L)$ in the self-dual case  \cite{IkJS12}
\begin{equation}
\label{logamp}
  \kappa(L) = \frac{\pi^2\gamma}{8(\pi-2\gamma)}\,\frac{1}{\log L^2}\,,
\end{equation}
valid for anisotropies $0\le\gamma<\pi/2$.\footnote{%
  For $\gamma=\pi/2$ the model has an $OSP(2|2)$ symmetry \cite{GaMa06}.
  The coupling constant can be obtained from RG arguments which determine its
  system size dependence at intermediate scales, giving $\kappa(L) \simeq
  1/\log L$ \cite{IkJS08}.}
Numerical data for general staggering suggested that the finite-size spectrum
is in fact independent of $\alpha\in (\gamma,\pi/2]$ and consistent with this
expression \cite{FrMa12}.

To go further, Candu and Ikhlef \cite{CaIk13} have reformulated the spectral
problem for the self-dual model in
terms of nonlinear integral equations.  In the next section we shall
derive such integral equations which are valid for arbitrary staggering
$\gamma<\alpha\le \pi/2$ and which are then used to analyze the continuous 
part of the spectrum and the corrections to scaling $R(L)$ in (\ref{fse}).

%%%%%%%%%%%%%%%%%%%%%%%%%%%%%%%%%%%%%%%%%%%%%%%%%%%%%%%%%%%%%%%%%%%%%%
\section{Nonlinear Integral Equations}
\begin{figure}[t]
\begin{center}
\includegraphics[height=7cm]{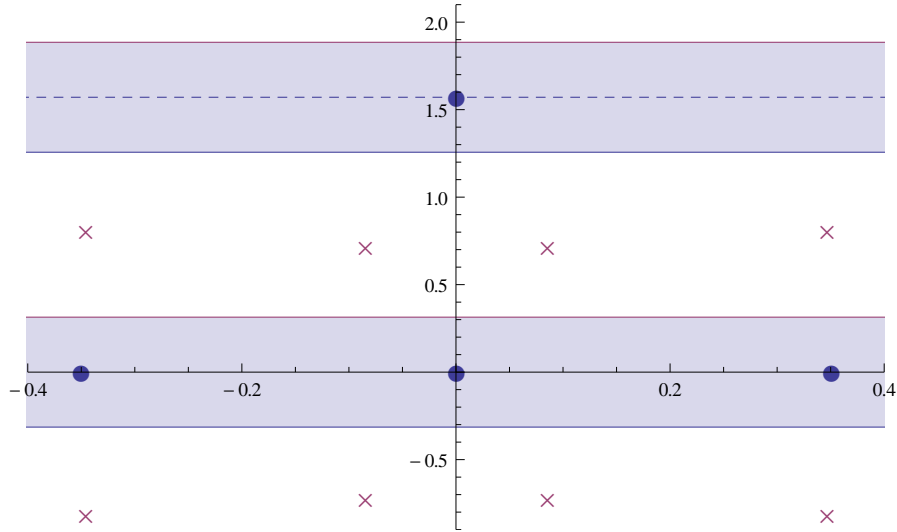}
\end{center}
\caption{$n_1=3$ and $n_2=1$ Bethe roots ($\bullet$) for $\gamma=\frac\pi5$
  and $\alpha=\frac{2\pi}{5}$ in the sector $L=n_1+n_2=4$. The $2L=8$
  hole-type solutions ($\times$), additional solutions to (\ref{BAequ})
  for fixed $\{\lambda_\ell\}_{l=1}^M$,
  are outside the shaded strips $|\Im
  z|\leq\frac{\gamma}{2}$ and $|\Im z-\frac{\pi}{2}|\leq\frac{\gamma}{2}$ of
  the complex plane.}
\label{rootdistribution}
\end{figure}

\begin{figure}[t]
\begin{center}
\includegraphics[height=6cm]{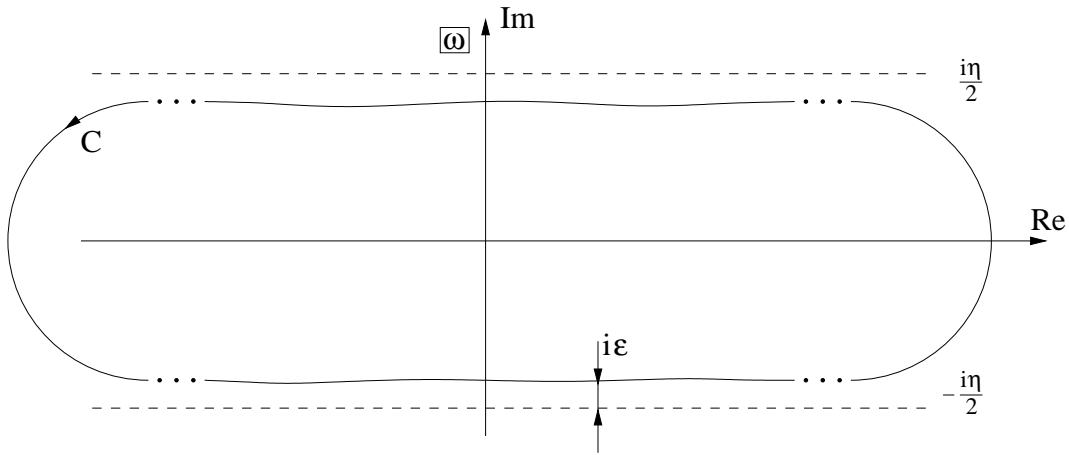}
\end{center}
\caption{The closed contour $\mathcal{C}$ in the complex plane: the parameter
  $\eta\equiv\frac12\min\big\{\gamma,\alpha-\gamma\big\}$
  determines the height of both shaded strips in Fig.~\ref{rootdistribution}
  separating the Bethe roots from the hole-type solutions and
  $\alpha$-dependent singularities of the auxiliary functions.  Evaluating
  $\mathfrak{a}_{1}$ for $\lambda\in\mathcal{C}$ by enclosing the real Bethe
  numbers $\{\nu_j\}_{j=1}^{n_1}$ (similarly $\{\mu_j\}_{j=1}^{n_2}$
  for $\mathfrak{a}_{2}$) one has to introduce a distance parameter
  $\varepsilon>0$ to avoid singularities at the border $\Im\lambda=\pm\eta/2$,
  e.g.\ from the kernels. Extending $\mathcal{C}$ to $\pm\infty$ covers all
  system sizes $L$.}
%  To evaluate
%  the auxiliary functions for $\lambda\in\mathcal{C}$ one has stay away from
%  the singularities of the kernels and driving terms by introducing a distance
%  parameter $\varepsilon>0$ to
%  $\eta\equiv\min\big\{\frac\gamma2,\frac\alpha2-\frac\gamma2\big\}$. The
%  value of $\eta$ is such that $\mathcal{C}$ fits into in the shaded strips of
%  Fig.~\ref{rootdistribution} enclosing all Bethe numbers
%  $\{\nu_j\}_{j=1}^{n_1}$ and $\{\mu_j\}_{j=1}^{n_2}$ respectively. Extending
%  $\mathcal{C}$ to $\pm\infty$ covers all system sizes $L$.}
\label{ccont}
\end{figure}

%%%%%%%%%%%%%%%%%%%%%%%%%%%%%%%%%%%%%%%%%%%%%%%%%%%%%%%%%%%%%%%%%%%%%%
%
To be specific we consider solutions to the Bethe equations corresponding to
low energy excitations (\ref{fse}) with $(m,w)=(0,0)$.  These configurations
contain a total of $M=n_1+n_2\equiv L$ roots (\ref{Bethe-roots}) distributed
symmetrically around the imaginary axis
\begin{gather} 
\label{Bethe-roots-sym}
  \{\nu_j\}_{j=1}^{n_1}=\{-\nu_j\}_{j=1}^{n_1}\,,\quad
  \{\mu_j\}_{j=1}^{n_2}=\{-\mu_j\}_{j=1}^{n_2}\,,
\end{gather}
see Fig.~\ref{rootdistribution} for an example.
%======================================================================%
%======================================================================%
%
Following Refs.~\onlinecite{Kluemper92,Kluemper93} we introduce auxiliary
functions
\begin{equation} \label{auxf} \mathfrak{a}_1(\lambda)=\mathfrak{a}(\lambda)
  \equiv 
  d(\lambda) \prod_{\ell=1}^L\frac{\sh(\lambda-\lambda_\ell+\I\gamma)}{
    \sh(\lambda-\lambda_\ell-\I\gamma)} \,
\end{equation}
and $\mathfrak{a}_2(\lambda)\equiv \mathfrak{a}_1(\lambda+\I\pi/2)$, thereby
encoding the Bethe roots (\ref{Bethe-roots}) 
%$\nu_j$ and $\mu_j$
in the zeroes of $(1+\mathfrak{a}_1)(\lambda)$ and
$(1+\mathfrak{a}_2)(\lambda)$, respectively.  The additional zeroes of these
expressions are called hole-type solutions, c.f.\ Fig.~\ref{rootdistribution}.
This allows to rewrite the Bethe equations \eqref{BAequ} in the sector of
$n_1+n_2=M=L$ roots and parameter ranges $0<\gamma<\alpha\leq\pi/2$ in terms
of coupled nonlinear integral equations (NLIEs)
\begin{align}
\begin{aligned}\label{aux1}
\log\mathfrak{a}_1(\lambda) = &2\I\gamma L + 
L\log\left(\frac{\sh(\lambda+\frac{\I\alpha}2-\frac{\I\gamma}2)}
  {\sh(\lambda+\frac{\I\alpha}2+\frac{\I\gamma}2)}
\frac{\sh(\lambda-\frac{\I\alpha}2-\frac{\I\gamma}2)}
  {\sh(\lambda-\frac{\I\alpha}2+\frac{\I\gamma}2)}\right)\\
&-\int_\mathcal{C}\frac{\D \omega}{2\pi}
  K_{\I\gamma}(\lambda-\omega)\log(1+\mathfrak{a}_1)(\omega)
+\int_\mathcal{C}\frac{\D \omega}{2\pi} 
  K_{\I\gamma^\prime}(\lambda-\omega)\log(1+\mathfrak{a}_2)(\omega)\,,
\end{aligned}\\[2ex]
\begin{aligned}\label{aux2}
\log\mathfrak{a}_2(\lambda) = &2\I\gamma L +  
L\log\left(\frac{\ch(\lambda+\frac{\I\alpha}2-\frac{\I\gamma}2)}
  {\ch(\lambda+\frac{\I\alpha}2+\frac{\I\gamma}2)}
\frac{\ch(\lambda-\frac{\I\alpha}2-\frac{\I\gamma}2)}
  {\ch(\lambda-\frac{\I\alpha}2+\frac{\I\gamma}2)}\right)\\
&-\int_\mathcal{C}\frac{\D \omega}{2\pi}
  K_{\I\gamma}(\lambda-\omega)\log(1+\mathfrak{a}_2)(\omega)
+\int_\mathcal{C}\frac{\D \omega}{2\pi}
  K_{\I\gamma^\prime}(\lambda-\omega)\log(1+\mathfrak{a}_1)(\omega)\,.
\end{aligned}
\end{align}
Here we have chosen the branch cut of the logarithm along the negative real
axis.  
Note that unlike in the NLIEs derived earlier for the self-dual model
\cite{CaIk13} the convolution integrals in (\ref{aux1}) and (\ref{aux2}) are
computed along a fixed non-intersecting closed contour~$\mathcal{C}$ (c.f.\
Fig.~\ref{ccont}).  As a consequence, the system size $L$ enters only as a
parameter in the driving terms.  The kernels
\begin{equation}
K_{\I\gamma}(x) = 
\frac{1}{\I}\frac{\sh(2\I\gamma)}{\sh(x-\I\gamma)\sh(x+\I\gamma)}
\,,\quad \gamma^\prime\equiv  \frac{\pi}{2}-\gamma 
\end{equation}
have no poles which need to be considered in the contour integration.
Furthermore, their Fourier transformations on the contour are regular which is
particularly useful for the numerical evaluation of the convolution integrals.

The Bethe numbers and thus the auxiliary functions $\mathfrak{a}_1$ and 
$\mathfrak{a}_2$ fix the eigenvalue of the transfer matrix. For $\lambda$ 
inside the closed contour $\mathcal{C}$ this yields
\begin{equation} \label{eigenvalue-integral}
\log\Lambda(\lambda) = 
-\I\gamma L + \int_\mathcal{C}\frac{\D \omega}{2\pi\I}
\frac{\sh(\I\gamma)\log(1+\mathfrak{a}_1)(\omega)}
  {\sh(\lambda-\omega)\sh(\lambda-\omega-\I\gamma)}
- \int_\mathcal{C}\frac{\D \omega}{2\pi\I}
\frac{\sh(\I\gamma)\log(1+\mathfrak{a}_2)(\omega)}
  {\ch(\lambda-\omega)\ch(\lambda-\omega-\I\gamma)}\,.
\end{equation}
%
%======================================================================%
%======================================================================%

The proof is straightforward and only involves Cauchy's theorem. 
Considering the logarithmic derivative of $\mathfrak{a}(\lambda)$,
\begin{equation*}
\partial\log\frac{\mathfrak{a}(\lambda)}{d(\lambda)} =
- \sum_{\ell=1}^{n_1} \frac{\sh(2\I\gamma)}
   {\sh(\lambda-\nu_\ell+\I\gamma)\sh(\lambda-\nu_\ell-\I\gamma)}
+ \sum_{\ell=1}^{n_2} \frac{\sh(2\I\gamma)}
   {\ch(\lambda-\mu_\ell+\I\gamma)\ch(\lambda-\mu_\ell-\I\gamma)}
\end{equation*}
the summation part can be cast into an integral representation involving the
auxiliary functions: for any analytic $f(\lambda)$ the relations
\begin{equation} \label{residuethm}
\sum_{\ell=1}^{n_1} f(\nu_\ell) = \int_\mathcal{C}\frac{\D \omega}{2\pi\I} 
  \frac{f(\omega)\partial\mathfrak{a}_1(\omega)}{1+\mathfrak{a}_1(\omega)} 
  \quad , \quad
\sum_{\ell=1}^{n_2} f(\mu_\ell) = \int_\mathcal{C}\frac{\D \omega}{2\pi\I} 
  \frac{f(\omega)\partial\mathfrak{a}_2(\omega)}{1+\mathfrak{a}_2(\omega)}
\end{equation}
hold if the contour encloses all Bethe roots but none of the hole-type
solutions and $\alpha$-dependent singularities of 
$1+\mathfrak{a}_{1,2}(\lambda)=0$. For the example shown in
Fig.~\ref{rootdistribution} this is guaranteed for the contour~$\mathcal{C}$
according to Fig.~\ref{ccont}.
%enclosing the axes $\mathrm{Im}(\lambda)=0$ and $\mathrm{Im}(\lambda)=\pi/2$ 
%which are located in the shaded strips.
%
As the expression $d(\lambda)$ is analytic within~$\mathcal{C}$
Eq.~(\ref{aux1}) immediately follows, where the integration constant can be
fixed by considering \eqref{auxf} for e.g.\
$\lambda\to\infty$. Eq.~(\ref{aux2}) is just a shift in the argument by
$\I\pi/2$ formally exchanging $\mathfrak{a}_1
\leftrightarrow\mathfrak{a}_2$. The eigenvalue $\Lambda(\lambda)$ can be
treated in a similar way: by factorizing the eigenvalue
\eqref{basiceigenvalue} in favour of the auxiliary function
$1+\mathfrak{a}_1(\lambda)$ and taking the logarithmic derivative,
\begin{equation*}
\partial\log\Lambda(\lambda) = 
\sum_{\ell=1}^{n_1}\frac{\sh(\I\gamma)}
  {\sh(\lambda-\nu_\ell)\sh(\lambda-\nu_\ell-\I\gamma)} 
-\sum_{\ell=1}^{n_2}\frac{\sh(\I\gamma)}
  {\ch(\lambda-\mu_\ell)\ch(\lambda-\mu_\ell-\I\gamma)} 
+\partial\log\big[1+\mathfrak{a}_1(\lambda)\big]\,,
\end{equation*}
the summation part can be cast into an integral representation according to 
\eqref{residuethm} as long as $\lambda$ and $\lambda-\I\gamma$ remain outside 
the closed contour. For $\lambda$ inside the closed contour $\mathcal{C}$ the 
additional term $\partial\log\big[1+\mathfrak{a}_1(\lambda)\big]$
can be absorbed into the contour integral involving $\mathfrak{a}_1$ by
Cauchy's theorem. Lifting the derivatives immediately yield 
(\ref{eigenvalue-integral}).

%%%%%%%%%%%%%%%%%%%%%%%%%%%%%%%%%%%%%%%%%%%%%%%%%%%%%%%%%%%%%%%%%%%%%%%

\subsection{Mixed Eigenvalues}
The energy $E$ and quasi momentum $K$ of the eigenstates are related to the
logarithmically combined expressions \eqref{finaleqs} evaluated at the points
$\xi$ and $\xi+\I\alpha$ being zeroes of $\mathfrak{a}(\lambda)$.  The bulk
parts can be split off by considering the mixed eigenvalues
\begin{equation}\label{mixed}
\Lambdamix(\lambda)\equiv
\Lambda(\lambda)\Lambda\Big(\lambda+\frac{\I\pi}2\Big) 
  \quad ,\quad
\Omegamix(\lambda) \equiv
\frac{\Lambda(\lambda)}{\Lambda\big(\lambda+\frac{\I\pi}2\big)}\;.
\end{equation}
Due to $\I\pi$-periodicity of \eqref{basiceigenvalue} the mixed eigenvalues
are periodic, $\log\Lambdamix(\lambda+{\I\pi}/{2})=\log\Lambdamix(\lambda)$,
and antiperiodic, $\log\Omegamix(\lambda+{\I\pi}/{2})=-\log\Omegamix(\lambda)$
with respect to $\I\pi/2$ satisfying the functional equations
\begin{multline}\label{funkE}
\Lambdamix\big(x-\frac{\I\gamma}{2}\big) 
\Lambdamix\big(x+\frac{\I\gamma}{2}\big)=
d\big(x-\frac{\I\gamma}{2}\big) 
d\big(x-\frac{\I\gamma}{2}+\frac{\I\pi}{2}\big)\times\\
\times\Big[1+\mathfrak{a}^{-1}\big(x-\frac{\I\gamma}{2}\big)\Big]
\Big[1+\mathfrak{a}^{-1}\big(x+\frac{\I\pi}{2}-\frac{\I\gamma}{2}\big)\Big]
\Big[1+\mathfrak{a}\big(x-\frac{\I\pi}{2}+\frac{\I\gamma}{2}\big)\Big]
\Big[1+\mathfrak{a}\big(x+\frac{\I\gamma}{2}\big)\Big]
\end{multline}
and
\begin{multline}\label{funkK}
\Omegamix\big(x-\frac{\I\gamma}{2}\big) 
\Omegamix\big(x+\frac{\I\gamma}{2}\big)
=\frac{d(x-\frac{\I\gamma}{2})}{d(x-\frac{\I\gamma}{2}+\frac{\I\pi}{2})} 
\times\\
\times\frac{\big[1+\mathfrak{a}^{-1}\big(x-\frac{\I\gamma}{2}\big)\big]
\big[1+\mathfrak{a}\big(x+\frac{\I\gamma}{2}\big)\big]}
{\big[1+\mathfrak{a}^{-1}\big(x+\frac{\I\pi}{2}-\frac{\I\gamma}{2}\big)\big]
\big[1+\mathfrak{a}\big(x-\frac{\I\pi}{2}+\frac{\I\gamma}{2}\big)\big]}\,,
\end{multline}
respectively. To finally evaluate \eqref{finaleqs} for general staggering 
the equations \eqref{funkE} and \eqref{funkK} can be solved in Fourier space 
for $\Lambdamix$ and $\Omegamix$ using the transformation pair
\begin{equation}
\widehat{f}(k) = 
\int\displaylimits_{-\infty}^\infty\!\!\!\D x \,\E^{-\I k x} f(x) \, , \quad
f(x) = \int\displaylimits_{-\infty}^\infty \!\!\!\frac{\D k}{2\pi} \E^{\I k x} 
\widehat{f}(k) \, .
\end{equation}

As $\Lambdamix(\lambda)$ and $\Omegamix(\lambda)$ are analytic\footnote{valid
  for the range $0<\gamma\leq\frac{\alpha}{2}$; similar results can be
  obtained for $\frac{\alpha}{2}<\gamma<\alpha\leq\frac{\pi}{2}$} in
$\big\{\lambda\in\mathbb{C}\big|
\frac{\pi+2\gamma}{4}<\Im\lambda<\frac{\pi-\gamma}{2}\big\}$ and the region
enclosed by the contour $\mathcal{C}$ a standard manipulation in Fourier space
yields with the ${\I\pi}/{2}$-(anti)periodicity
\begin{align}\label{system1}
&\begin{aligned}
\partial\log\Lambdamix\big(x-\frac{\I\pi}{4}+&\frac{\I\gamma}{2}\big) 
= -\I L \int\displaylimits_{-\infty}^{\infty}\!\!
\frac{\D k\,\E^{\I k x}\sh\big(\frac{k\gamma}{2}\big)
  \ch\big(\frac{k\pi}{4}-\frac{k\alpha}{2}\big)}
{\sh\big(\frac{k\pi}{4}\big)\ch\big(\frac{k\pi}{4}-\frac{k\gamma}{2}\big)}\\
&-\sum_{j=1}^2\int\displaylimits_{-\infty}^\infty\!\!\frac{\D k}{2\pi}
\frac{\E^{\I k x}\,\E^{-\frac{k\gamma}{2}+k\varepsilon}}
  {2\ch\big(\frac{k\pi}{4}-\frac{k\gamma}{2}\big)}\,
\int\displaylimits_{-\infty}^\infty\!\!\!\D y \,\E^{-\I k y}
  \partial\log\big(1+\mathfrak{a}_j\big)
  \big(y-\frac{\I\gamma}{2}+\I\varepsilon\big)\\
&+\sum_{j=1}^2\int\displaylimits_{-\infty}^\infty\!\!\frac{\D k}{2\pi}
\frac{\E^{\I k x}\,\E^{\frac{k\gamma}{2}-k\varepsilon}}
  {2\ch\big(\frac{k\pi}{4}-\frac{k\gamma}{2}\big)}\,
\int\displaylimits_{-\infty}^\infty\!\!\!\D y \,\E^{-\I k y}
  \partial\log\big(1+\mathfrak{a}_j^{-1}\big)
  \big(y+\frac{\I\gamma}{2}-\I\varepsilon\big)\,,
\end{aligned}\\
&\begin{aligned}
\partial\log\Omegamix\big(x-\frac{\I\pi}{4}+&\frac{\I\gamma}{2}\big) 
= \;\I L \int\displaylimits_{-\infty}^{\infty}\!\!
\frac{\D k\,\E^{\I k x}\sh\big(\frac{k\gamma}{2}\big)
  \sh\big(\frac{k\pi}{4}-\frac{k\alpha}{2}\big)}
{\ch\big(\frac{k\pi}{4}\big)\sh\big(\frac{k\pi}{4}-\frac{k\gamma}{2}\big)}\\
&+\int\displaylimits_{-\infty}^\infty\!\!\frac{\D k}{2\pi}
\frac{\E^{\I k x}\,\E^{-\frac{k\gamma}{2}+k\varepsilon}}
{2\sh\big(\frac{k\pi}{4}-\frac{k\gamma}{2}\big)}\,
\int\displaylimits_{-\infty}^\infty\!\!\!\D y \,\E^{-\I k y}\partial\log 
\frac{1+\mathfrak{a}_1\big(y-\frac{\I\gamma}{2}+\I\varepsilon\big)}
  {1+\mathfrak{a}_2\big(y-\frac{\I\gamma}{2}+\I\varepsilon\big)}\\
&-\int\displaylimits_{-\infty}^\infty\!\!\frac{\D k}{2\pi}
\frac{\E^{\I k x}\,\E^{\frac{k\gamma}{2}-k\varepsilon}}
  {2\sh\big(\frac{k\pi}{4}-\frac{k\gamma}{2}\big)}\,
\int\displaylimits_{-\infty}^\infty\!\!\!\D y \,\E^{-\I k y}\partial\log 
\frac{1+\mathfrak{a}_1^{-1}\big(y+\frac{\I\gamma}{2}-\I\varepsilon\big)}
  {1+\mathfrak{a}_2^{-1}\big(y+\frac{\I\gamma}{2}-\I\varepsilon\big)}
\end{aligned}\label{system2}
\end{align}
where we used the Fourier transform of $\log d(\lambda)$,
\begin{equation}
\widehat{\partial\log d}(k)=
-\frac{4\pi\I L\sh({\frac{k\gamma}{2}})}
  {\sh(\frac{k\pi}{2})}\ch\Big(\frac{k\pi}{2}-\frac{k\alpha}{2}\Big) \quad .
\end{equation}

Note that this system (\ref{system1}) and (\ref{system2}) already describes
the energy~$E$ and quasi momentum~$K$ according to \eqref{finaleqs} in the 
self dual case $\alpha=\pi/2$. However, 
as $\Lambdamix(\lambda)$ and $\Omegamix(\lambda)$ are composed from 
simple eigenvalues $\Lambda(\lambda)$, c.f.\ \eqref{mixed}, one can solve 
the system for 
$\Lambda\big(x-\frac{\I\pi}{4}+\frac{\I\gamma}{2}\big)$ 
and $\Lambda\big(x+\frac{\I\pi}{4}+\frac{\I\gamma}{2}\big)$. Recombining 
after suitably shifting the arguments \eqref{finaleqs} reads in Fourier 
representation
\begin{align}
&\begin{aligned}
\partial\log\Lambda(&x+\xi)+\partial\log\Lambda(x+\xi+\I\alpha) =\\
= &-\I L \int\displaylimits_{-\infty}^{\infty}\!\!
\frac{\D k\,\E^{\I k x}\sh\big(\frac{k\gamma}{2}\big)
  \ch^2\big(\frac{k\pi}{4}-\frac{k\alpha}{2}\big)}
{\sh\big(\frac{k\pi}{4}\big)\ch\big(\frac{k\pi}{4}-\frac{k\gamma}{2}\big)}
-\I L \int\displaylimits_{-\infty}^{\infty}\!\!
\frac{\D k\,\E^{\I k x}\sh\big(\frac{k\gamma}{2}\big)
  \sh^2\big(\frac{k\pi}{4}-\frac{k\alpha}{2}\big)}
{\ch\big(\frac{k\pi}{4}\big)\sh\big(\frac{k\pi}{4}-\frac{k\gamma}{2}\big)}\\
&-\sum_{j=1}^2\int\displaylimits_{-\infty}^\infty\!\!\frac{\D k}{2\pi}
\frac{\ch\big(\frac{k\pi}{4}-\frac{k\alpha}{2}\big)}
  {2\ch\big(\frac{k\pi}{4}-\frac{k\gamma}{2}\big)}\,
{\E^{\I k x}\,\E^{-\frac{k\gamma}{2}+k\varepsilon}}\,
\int\displaylimits_{-\infty}^\infty\!\!\!\D y \,\E^{-\I k y}
  \partial\log\big(1+\mathfrak{a}_j\big)
  \big(y-\frac{\I\gamma}{2}+\I\varepsilon\big)\\
&+\sum_{j=1}^2\int\displaylimits_{-\infty}^\infty\!\!\frac{\D k}{2\pi}
\frac{\ch\big(\frac{k\pi}{4}-\frac{k\alpha}{2}\big)}
  {2\ch\big(\frac{k\pi}{4}-\frac{k\gamma}{2}\big)}\,
{\E^{\I k x}\,\E^{\frac{k\gamma}{2}-k\varepsilon}}\,
\int\displaylimits_{-\infty}^\infty\!\!\!\D y \,\E^{-\I k y}
  \partial\log\big(1+\mathfrak{a}^{-1}_j\big)
  \big(y+\frac{\I\gamma}{2}-\I\varepsilon\big)\\
&-\int\displaylimits_{-\infty}^\infty\!\!\frac{\D k}{2\pi}
\frac{\sh\big(\frac{k\pi}{4}-\frac{k\alpha}{2}\big)}
  {2\sh\big(\frac{k\pi}{4}-\frac{k\gamma}{2}\big)}\,
{\E^{\I k x}\,\E^{-\frac{k\gamma}{2}+k\varepsilon}}\,
\int\displaylimits_{-\infty}^\infty\!\!\!\D y \,\E^{-\I k y}
  \partial\log\bigg(\frac{1+\mathfrak{a}_1}{1+\mathfrak{a}_2}\bigg)
  \big(y-\frac{\I\gamma}{2}+\I\varepsilon\big)\\
&+\int\displaylimits_{-\infty}^\infty\!\!\frac{\D k}{2\pi}
\frac{\sh\big(\frac{k\pi}{4}-\frac{k\alpha}{2}\big)}
  {2\sh\big(\frac{k\pi}{4}-\frac{k\gamma}{2}\big)}\,
{\E^{\I k x}\,\E^{+\frac{k\gamma}{2}-k\varepsilon}}\,
\int\displaylimits_{-\infty}^\infty\!\!\!\D y \,\E^{-\I k y}
  \partial\log\bigg(\frac{1+\mathfrak{a}^{-1}_1}{1+\mathfrak{a}^{-1}_2}\bigg)
  \big(y+\frac{\I\gamma}{2}-\I\varepsilon\big)\,,\\
\end{aligned}\\[1em]
&\begin{aligned}
\partial\log\Lambda(&x+\xi)-\partial\log\Lambda(x+\xi+\I\alpha) =\\
=& \;\I L \int\displaylimits_{-\infty}^{\infty}\!\!
\frac{\D k\,\E^{\I k x}\sh\big(\frac{k\gamma}{2}\big)
  \sh\big(\frac{k\pi}{2}-k\alpha\big)}
{2\ch\big(\frac{k\pi}{4}\big)\sh\big(\frac{k\pi}{4}-\frac{k\gamma}{2}\big)}
+ \I L \int\displaylimits_{-\infty}^{\infty}\!\!
\frac{\D k\,\E^{\I k x}\sh\big(\frac{k\gamma}{2}\big)
  \sh\big(\frac{k\pi}{2}-k\alpha\big)}
{2\sh\big(\frac{k\pi}{4}\big)\ch\big(\frac{k\pi}{4}-\frac{k\gamma}{2}\big)}\\
&+\sum_{j=1}^2\int\displaylimits_{-\infty}^\infty\!\!\frac{\D k}{2\pi}
\frac{\sh\big(\frac{k\pi}{4}-\frac{k\alpha}{2}\big)}
  {2\ch\big(\frac{k\pi}{4}-\frac{k\gamma}{2}\big)}\,
{\E^{\I k x}\,\E^{-\frac{k\gamma}{2}+k\varepsilon}}\,
\int\displaylimits_{-\infty}^\infty\!\!\!\D y \,\E^{-\I k y}
  \partial\log\big(1+\mathfrak{a}_j\big)
  \big(y-\frac{\I\gamma}{2}+\I\varepsilon\big)\\
&-\sum_{j=1}^2\int\displaylimits_{-\infty}^\infty\!\!\frac{\D k}{2\pi}
\frac{\sh\big(\frac{k\pi}{4}-\frac{k\alpha}{2}\big)}
  {2\ch\big(\frac{k\pi}{4}-\frac{k\gamma}{2}\big)}\,
{\E^{\I k x}\,\E^{\frac{k\gamma}{2}-k\varepsilon}}\,
\int\displaylimits_{-\infty}^\infty\!\!\!\D y \,\E^{-\I k y}
  \partial\log\big(1+\mathfrak{a}^{-1}_j\big)
  \big(y+\frac{\I\gamma}{2}-\I\varepsilon\big)\\
&+\int\displaylimits_{-\infty}^\infty\!\!\frac{\D k}{2\pi}
\frac{\ch\big(\frac{k\pi}{4}-\frac{k\alpha}{2}\big)}
  {2\sh\big(\frac{k\pi}{4}-\frac{k\gamma}{2}\big)}\,
{\E^{\I k x}\,\E^{-\frac{k\gamma}{2}+k\varepsilon}}\,
\int\displaylimits_{-\infty}^\infty\!\!\!\D y \,\E^{-\I k y}
  \partial\log\bigg(\frac{1+\mathfrak{a}_1}{1+\mathfrak{a}_2}\bigg)
\big(y-\frac{\I\gamma}{2}+\I\varepsilon\big)\\
&-\int\displaylimits_{-\infty}^\infty\!\!\frac{\D k}{2\pi}
\frac{\ch\big(\frac{k\pi}{4}-\frac{k\alpha}{2}\big)}
  {2\sh\big(\frac{k\pi}{4}-\frac{k\gamma}{2}\big)}\,
{\E^{\I k x}\,\E^{+\frac{k\gamma}{2}-k\varepsilon}}\,
\int\displaylimits_{-\infty}^\infty\!\!\!\D y \,\E^{-\I k y}
  \partial\log\bigg(\frac{1+\mathfrak{a}^{-1}_1}{1+\mathfrak{a}^{-1}_2}\bigg)
  \big(y+\frac{\I\gamma}{2}-\I\varepsilon\big)\\
\end{aligned}
\end{align}
for general staggering $\gamma<\alpha\leq\pi/2$. From the bulk parts of 
these expressions we can read off the energy density 
$\varepsilon_\infty$ (\ref{ebulk}) 
and quasi momentum densities $k_\infty$ (\ref{kbulk}) already obtained 
within the root density approach above.

Using the relation $\mathfrak{a}^{-1}(-\lambda)=\mathfrak{a}(\lambda)$ 
provided by the Bethe root's symmetry \eqref{Bethe-roots-sym} energy 
and momentum reduce for all $0<\gamma<\alpha\leq\pi/2$ to
\begin{align}
&\begin{aligned}\label{energy}
E-L\varepsilon_\infty =\;&\I\hspace*{-.4em}
  \int\displaylimits_{-\infty}^\infty\!\!\frac{\D k}{2\pi}
\frac{\ch\big(\frac{k\pi}{4}-\frac{k\alpha}{2}\big)}
  {\ch\big(\frac{k\pi}{4}-\frac{k\gamma}{2}\big)}\,
{\E^{-\frac{k\eta}{2}+k\varepsilon}}\,\sum_{j=1}^2
\int\displaylimits_{-\infty}^\infty\!\!\!\D y \,\E^{-\I k y}
  \partial\log\big(1+\mathfrak{a}_j\big)
  \big(y-\frac{\I\eta}{2}+\I\varepsilon\big)\\
&+\I\hspace*{-.4em}\int\displaylimits_{-\infty}^\infty\!\!\frac{\D k}{2\pi}
\frac{\sh\big(\frac{k\pi}{4}-\frac{k\alpha}{2}\big)}
  {\sh\big(\frac{k\pi}{4}-\frac{k\gamma}{2}\big)}\,
{\E^{-\frac{k\eta}{2}+k\varepsilon}}\,
\int\displaylimits_{-\infty}^\infty\!\!\!\D y \,\E^{-\I k y}
  \partial\log\bigg(\frac{1+\mathfrak{a}_1}{1+\mathfrak{a}_2}\bigg)
\big(y-\frac{\I\eta}{2}+\I\varepsilon\big)\,,\\
\end{aligned}\\[1em]
&\begin{aligned} \label{momentum}
K-Lk_\infty =&\int\displaylimits_{-\infty}^\infty\!\!\frac{\D k}{2\pi}
\frac{\sh\big(\frac{k\pi}{4}-\frac{k\alpha}{2}\big)}
  {\ch\big(\frac{k\pi}{4}-\frac{k\gamma}{2}\big)}\,
{\E^{-\frac{k\eta}{2}+k\varepsilon}}\,\sum_{j=1}^2
\int\displaylimits_{-\infty}^\infty\!\!\!\D y \,\E^{-\I k y}
  \log\big(1+\mathfrak{a}_j\big)\big(y-\frac{\I\eta}{2}+\I\varepsilon\big)\\
&+\int\displaylimits_{-\infty}^\infty\!\!\frac{\D k}{2\pi}
\frac{\ch\big(\frac{k\pi}{4}-\frac{k\alpha}{2}\big)}
  {\sh\big(\frac{k\pi}{4}-\frac{k\gamma}{2}\big)}\,
{\E^{-\frac{k\eta}{2}+k\varepsilon}}\,
\int\displaylimits_{-\infty}^\infty\!\!\!\D y \,\E^{-\I k y}
  \log\bigg(\frac{1+\mathfrak{a}_1}{1+\mathfrak{a}_2}\bigg)
\big(y-\frac{\I\eta}{2}+\I\varepsilon\big)\,.\\
\end{aligned}
\end{align}
Note that the singularity from the kernel in the last line of \eqref{momentum}
is compensated by the zero of the auxiliary functions
\begin{equation}
\int\displaylimits_{-\infty}^\infty\!\!\!\D y \,
  \log\bigg(\frac{1+\mathfrak{a}_1}{1+\mathfrak{a}_2}\bigg)
\big(y-\frac{\I\eta}{2}+\I\varepsilon\big) =0 \quad .
\end{equation}

%%%%%%%%%%%%%%%%%%%%%%%%%%%%%%%%%%%%%%%%%%%%%%%%%%%%%%%%%%%%%%%%%%%%%%
\section{Analysis of the finite size spectrum}
%%%%%%%%%%%%%%%%%%%%%%%%%%%%%%%%%%%%%%%%%%%%%%%%%%%%%%%%%%%%%%%%%%%%%%
%
Based on the formulation of the eigenvalues of the Hamiltonian and the
quasi-momentum operator in terms of the solutions to the NLIEs (\ref{aux1}) and
(\ref{aux2}) we can now proceed with our analysis of the low energy spectrum
(\ref{fse}) of the staggered six-vertex model.  For this it will be particularly
important to vary the parameter $\widetilde{m}$ (\ref{BAnos}) in a controlled
way as the system size is increased.
As discussed in Sect.~\ref{sec:BAthermo} the density of the two types of Bethe
roots (\ref{Bethe-roots}) depends on the staggering parameter $\alpha$ in the
model, see Eq.~(\ref{TLstaggering}).  This implies that the ground state can
be realized only for certain commensurate system sizes $L=L_{\mathrm{c}}$: for
staggering
\begin{equation}
\label{ratstagg}
  \alpha=\alpha(p,q)\equiv \frac{p\gamma+q(\pi-\gamma)}{p+q} 
\end{equation}
with $p,q\in\mathbb{N}$ prime to each other ($p=q=1$ on the self-dual line) we
have to choose $L_{\mathrm{c}}$ being an integer multiple of $(p+q)$.
For this choice of parameters the ground state corresponding to $(m,w)=(0,0)$
and $\widetilde{m}=0$ is described by integer numbers $n_1^{(0)} = Lp/(p+q)$
and $n_2^{(0)}=Lq/(p+q)$ of Bethe roots on the real line and with
$\mathrm{Im}(\lambda_j)=\I\pi/2$, respectively.
Excitations can be constructed by shifting $\kah$ roots between
these two sets, i.e.\ with $n_{1,2} = n_{1,2}^{(0)}\pm \kah$, resulting in
$\widetilde{m}=2\kah$.  Similarly, we can consider the spectrum in sectors
where the commensurability condition is not satisfied: 
let $L=\ell_0(p+q) + r$ with $\ell_0\in\mathbb{N}$, $r=1,2,\ldots,(p+q-1)$.
The Bethe states are
described by $n_1=\ell_0 p+\kah+r$ and $n_2=\ell_0 q-\kah$ roots of the two 
types. From (\ref{BAnos}) we obtain $\widetilde{m}=2\kah+2rq/(p+q)$ for these
states. Together this allows to vary $\widetilde{m}$ in steps of $2q/(p+q)$ 
for the staggering parameter (\ref{ratstagg}).
%
%
% \textbf{For the numerical treatment we solve \eqref{aux1} and \eqref{aux2}
% by iteration. As our NLIEs are satisfied by all solutions from the sector
% $n_1+n_2=L$ the considered state with fixed $n_1$ and $n_2$ can be selected
% by the initial values for $\mathfrak{a}(\lambda)$, e.g.\ multiple zeroes
% $\nu_j=\mu_j=0$ in Eqs.~(\ref{Bethe-roots}) and (\ref{auxf}).}

The observed presence of a continuous component of the spectrum implies that
there should be a corresponding continuous quantum number in the thermodynamic
limit $L\to\infty$.  For the self-dual model, i.e.\ $\alpha=\pi/2$, it has
been shown that this quantum number is in fact related to the conserved
quasi-momentum $K$ of the corresponding excitation \cite{IkJS12,CaIk13}:
for large $L$ the number $\widetilde{m}$ characterizing the Bethe
configuration (\ref{BAnos}) is related to the rescaled quasi-momentum
$s \equiv \frac{\pi-2\gamma}{4\pi\gamma}\,K$
as
\begin{equation}
\label{mtvss}
  \widetilde{m} \simeq \frac{4s}{\pi}\, \left(\log\frac{L}{L_0} + B(s)\right)\,.
\end{equation}
Here $L_0$ is a non-universal length scale which only depends on the
anisotropy $\gamma$ while the function $B(s)$ determines the (finite part) of
the density of states in the continuum part of the spectrum.

\begin{figure}[t]
\begin{center}
\includegraphics[width=0.7\textwidth]{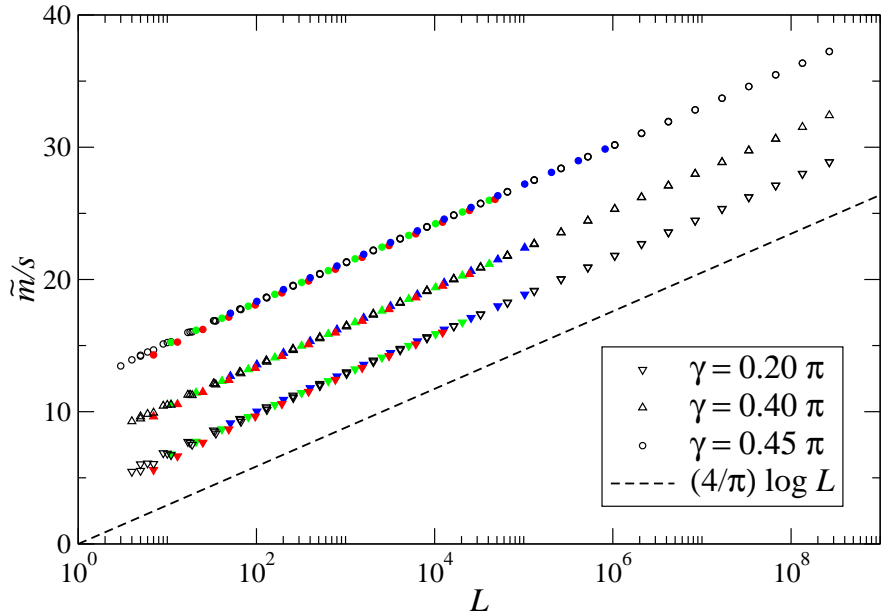}
\end{center}
\caption{Ratio of the number $\widetilde{m}$ characterizing the configuration
  of Bethe roots according to (\ref{BAnos}) and the quantum number $s$
  (\ref{qnocont}) related to the quasi-momentum of that state as a function of
  the system size. The plot shows data of various states for anisotropies 
  $\gamma/\pi=0.2$, $0.4$ and $0.45$ and different values of the staggering
  $\alpha(2,1)=\frac{1}{3}(\pi+\gamma)$ (red symbols),
  $\alpha(3,2)=\frac{1}{5}(2\pi+\gamma)$ (green symbols),
  $\alpha(13,12)=\frac{1}{25}(12\pi+\gamma)$ (blue symbols) and the self-dual
  case $\alpha(1,1)=\frac{\pi}{2}$ (black open symbols).  The linear
  dependence on $\log L$ (dashed line) with an offset depending only on
  $\gamma$ is clearly seen.}
\label{figqno}
\end{figure}

This line of arguments can be implemented in a straigthforward way for
staggering away from the self-dual line: here the quasi-momentum has a non
zero value (\ref{kbulk}) in the ground state ($(m,w)=(0,0)$ and
$\widetilde{m}=0$).  Since the expression (\ref{logamp}) has been found to be
consistent with numerical results \cite{FrMa12} for the finite size spectrum
for $\alpha\in (\gamma,\pi/2]$ we propose that the quasi-momentum should be
replaced by its deviation from the ground state value, i.e.\ $(K-Lk_\infty)$
in the relations above.  In particular, the quantum number for the continuous
component of the finite size spectrum appearing in (\ref{mtvss}) should read
\begin{equation}
\label{qnocont}
  s \equiv \frac{\pi-2\gamma}{4\pi\gamma}\,(K-Lk_\infty)\,.
\end{equation}
To support this conjecture we have computed the quasi-momentum
(\ref{momentum}) by numerical solution of the NLIEs (\ref{aux1}), (\ref{aux2}):
starting from appropriate initial values for the auxiliary functions
(\ref{auxf}), e.g.\ by choosing the $L=n_1+n_2$ Bethe roots
(\ref{Bethe-roots}) to be degenerate $\nu_j=\mu_j=0$, the NLIEs can be solved
by iteration.  In Fig.~\ref{figqno} we present data for the ratio
$\widetilde{m}/s$ obtained using this procedure as a function of the system
size for various values of the anisotropy $\gamma$ and the staggering
$\alpha$.  Apart from the corrections for small $L$ the numerical values show
the predicted linear dependence (\ref{mtvss}) on $\log L$ with an offset
independent of the staggering.
%
%======================================================================%
%======================================================================%

Rewriting the expression (\ref{fse}) for the finite size spectrum in terms of
the quantum number (\ref{qnocont}) the effective field theory describing the
low energy excitations of the lattice model can now be identified using the
predictions of conformal field theory (CFT) \cite{BlCN86,Affl86}
\begin{equation}
\label{fscft}
  \Delta E = \frac{2\pi v_F}{L} \left( -\frac{c}{12} + h + \bar{h} \right) \,,
  \quad
  P = \frac{2\pi}{L}\left(h-\bar{h}\right)\,.
\end{equation}
The finite size energy of the ground state ($(m,w)=(0,0)$ and
$\widetilde{m}=0$) implies that the effective central charge of the vertex
model is $c_{\mathrm{eff}}=2$, as expected for a model with two species of
excitations (holes in the distributions of Bethe roots with
$\mathrm{Im}(\lambda)=0$ or $\pi/2$, see Eq.~(\ref{Bethe-roots})).  On the
other hand, because of the presence of the continuous quantum number
(\ref{qnocont}), the continuum limit of the spin chain has to correspond to a
non-rational CFT with non-normalizable Virasoro vacuum and therefore a state
with $h=\bar{h}=0$ should not be part of the spectrum of the lattice model.
This has led Ikhlef \emph{et al.} \cite{IkJS08} to identify the continuum
limit of the staggered six-vertex model for $\alpha=\pi/2$ with the
$SL(2,\mathbb{R})/U(1)$ sigma model at level $k=\pi/\gamma\in(2,\infty)$
describing a two-dimensional Euclidean black hole \cite{Witten91}.  This CFT
has central charge $c = 2\,\frac{k+1}{k-2}$ and primary fields with dimensions
\begin{equation}
  \label{cftbh}
%  c = 2\,\frac{k+1}{k-2}\,, \quad
  h = \frac{(m-kw)^2}{4k} + \frac{s^2+1/4}{k-2}\,,\quad
  \bar{h} = \frac{(m+kw)^2}{4k} + \frac{s^2+1/4}{k-2}\,.
\end{equation}
Here, $j=(-1/2+\I s)$ with real $s$ is the spin of the $SL(2,\mathbb{R})$
affine primaries from the principal continuous representations.  
%
% In addition, there exist discrete representations with real spin
% $\frac{1-k}{2}<j<-\half$ and .
%
The ground state of the lattice model corresponds to the state with lowest
conformal weight $h_0=\bar{h}_0=1/(4(k-2))$ which results in the observed
effective central charge.
Further evidence for this proposal has been provided in
Refs.~\onlinecite{IkJS12,CaIk13} where the density of states in the continuum
has been computed from the finite part $\propto B(s)$ in the relation
(\ref{mtvss}) and shown to agree with the known result for the
$SL(2,\mathbb{R})/U(1)$ sigma model \cite{MaOS01,HaPT02}
\begin{equation}
\label{BBH}
\begin{aligned}
  \rho_{BH}(s) &= \frac{1}{\pi} \big( \log\epsilon + \partial_s (s\,B_{BH}(s))
  \big)\,, \\
  B_{BH}(s) &= \frac{1}{2s} \mathrm{Im} \log \left[ 
    \Gamma\left(\frac{1}{2}\,( 1-m+w\,k) -\I s\right)
    \Gamma\left(\frac{1}{2}\,( 1-m-w\,k) -\I s\right)\right]\,
\end{aligned}
\end{equation}
(the term $\log\epsilon$ arises from the regularization needed to handle
divergencies arising in the string theory).

From its effect on the finite size spectrum a staggering $\alpha\ne\pi/2$ is
an irrelevant perturbation of the low energy effective theory.  This is
consistent with the assumption
%Therefore we expect 
that the critical theory for the entire phase $0\le\gamma< \alpha <\pi-\gamma$
is the Euclidean black hole sigma model CFT.  Due to the presence of a
continuous spectrum, however, it is not sufficient for this identification to
rely on the finite size spectrum alone.  In addition the density of states in
the continuum has to be computed.
According to the considerations at the beginning of this section, the allowed
values of $\widetilde{m}$ for given staggering $\alpha$ and system size $L$
differ by multiples of $2$.  Therefore, the density of states in the continuum
follows from (\ref{mtvss}) to be (see also Refs.~\onlinecite{IkJS12,CaIk13})
\begin{equation}
  \rho(s) = \half\partial_s \widetilde{m} = \frac{2}{\pi}\left(
    \log\frac{L}{L_0} + \partial_s(sB(s))\right)\,.
\end{equation}

\begin{figure}[t]
\begin{center}
\includegraphics[width=0.7\textwidth]{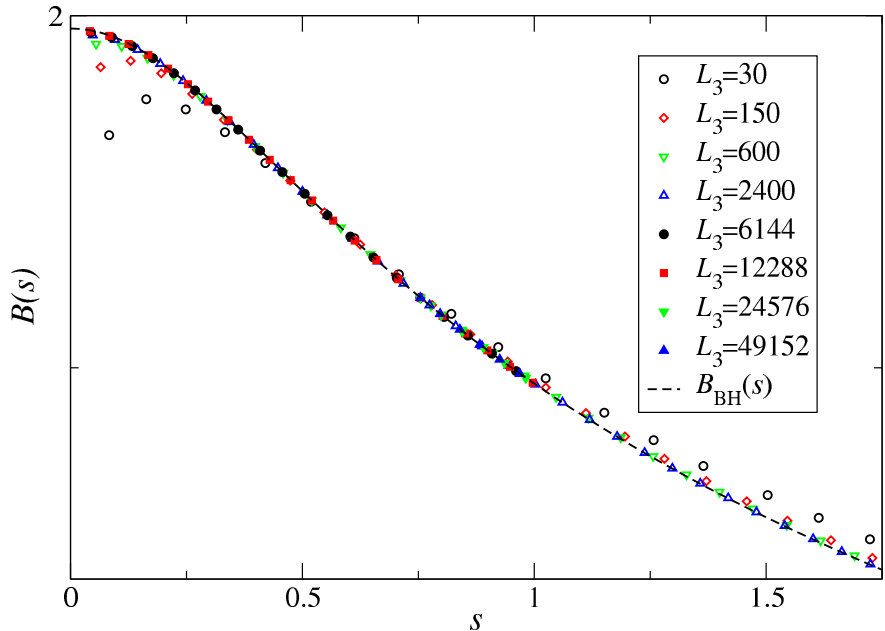}
\end{center}
\caption{Numerical values for the finite part $B(s)$ of the density of states
  in the continuous part of the spectrum of the staggered six-vertex model.
  Data are presented for system sizes $L=L_c+r$, $r=0,1,2$, at anisotropy
  $\gamma=\frac\pi5$ and staggering $\alpha(p=2,q=1)=\frac{1}{3}(\pi+\gamma)$.
  The dashed line is the result for the $SL(2,\mathbb{R})/U(1)$ sigma model,
  see Eq.~(\ref{BBH}).}
\label{figdos}
\end{figure}

%Using (\ref{mtvss}) w
We have computed the function $B(s)$ for staggering $\alpha\ne\pi/2$ by
numerical solution of the Bethe equations (\ref{BAequ}) for systems sizes up
to $L\lesssim 2400$ and from the NLIEs (\ref{aux1}), (\ref{aux2}) for larger
$L$.  By comparing the data for the largest system size available with the
corresponding quantity for the black hole sigma model (\ref{BBH}) we have
determined the non-universal length scale $L_0(\gamma)$.  As seen in
Fig.~\ref{figdos} the convergence of the data obtained for the lattice model
towards $B_{BH}(s)$ is excellent, just as in the previous studies for the
self-dual case \cite{IkJS12,CaIk13}.
%
%======================================================================%
%======================================================================%

%%%%%%%%%%%%%%%%%%%%%%%%%%%%%%%%%%%%%%%%%%%%%%%%%%%%%%%%%%%%%%%%%%%%%%
\section{Corrections to scaling}
%%%%%%%%%%%%%%%%%%%%%%%%%%%%%%%%%%%%%%%%%%%%%%%%%%%%%%%%%%%%%%%%%%%%%%
For the identification of the CFT describing the continuum limit of the
staggered six-vertex model in the previous section we have made use of the
consequences of conformal invariance on the finite size spectrum (\ref{fscft})
which hold asymptotically for $L\to\infty$.  Having revealed the origin of the
logarithmic terms in the spectrum to be the presence of a continuous spectrum
of critical exponents we now turn to analyze the dominant corrections to
finite size scaling (\ref{fse}) of an energy eigenvalue $E_a(L)$ corresponding
to the operator $\Phi_a$ in the CFT with conformal weights $(h_a,\bar{h}_a)$,
i.e.\
\begin{equation}\label{corrfsc}
  R_a(L) = \frac{L}{2\pi v_F} \left(E_a(L)-L\varepsilon_\infty\right)
       + \frac{c}{12} - (h_a+\bar{h}_a)\,.
\end{equation}
These corrections arise from deviations of the lattice Hamiltonian from the
fixed-point Hamiltonian $H^*$ of the CFT by terms involving irrelevant
operators \cite{Card86a} and therefore should provide additional insights 
into the particular lattice regularization of the CFT considered. If the
deviations are small these terms can be written as
\begin{equation}
  \label{cftpert}
  H_{\mathrm{lattice}} = H^* + \sum_b g_b \int\! \D x\,\Phi_b(x)\, 
\end{equation}
where $\Phi_{b}$ are conformal fields with scaling dimension
$X_b=h_b+\bar{h}_b>2$ and conformal spin $s_b=h_b-\bar{h}_b$.  The coupling
constants $g_b$ are in general unknown.

The effect of these terms on the finite size spectrum can be studied within
perturbation theory \cite{Card86a,AlBB88}: to second order one finds
\begin{equation}
\label{corrsc}
  R_a(L) \simeq 2\pi \sum_b g_b\, C_{a,a,b} \left(\frac{2\pi}{L}\right)^{X_b-2}
  + 4\pi^2\sum_{\substack{b,b',a'\\a'\ne a}} g_b g_{b'}\,
  \frac{C_{a,a',b}\,C_{a',a,b'}}{X_a-X_{a'}} 
  \left(\frac{2\pi}{L}\right)^{X_b+X_b'-4} +\ldots
\end{equation}
Here conformal invariance has been used to compute the matrix elements of the
perturbation (\ref{cftpert}) from the three point functions $\langle
\Phi_a(z_1) \Phi_b(z_2) \Phi_c(z_3) \rangle$ in the complex plane 
\begin{equation}
\label{corr3pt}
  \langle a | \Phi_b(x) | c \rangle = C_{a,b,c}
  \left(\frac{2\pi}{L}\right)^{X_b} \, \E^{\frac{2\pi\I}{L}(s_a-s_c)x}\,,
\end{equation}
$C_{a,b,c}$ are the universal coefficients appearing in the operator product
expansion (OPE) of primary fields in the CFT.
The first order term in (\ref{corrsc}) is absent for the ground state, an
exception being contributions due to descendents with dimension $X=4$ of the
identity operator which are expected to be present in any theory
\cite{Card86a,Card86c,Rein87}.  Based on these insights the irrelevant
operators present in lattice formulations of various unitary models have been
identified:
the 'analytic' corrections to scaling resulting from operators in the
conformal block of the identity have been studied to even higher orders as in
(\ref{corrsc}), c.f.\ \cite{GeRV87,Rein87,IzHU01}.  For the six-vertex model
including its higher spin variants subject to various boundary conditions some
of the deviations from the respective fixed point have been identified based
on numerical studies of the finite size spectrum \cite{AlBB88,AlMa89,NiWF09}.
Very recently, it has been shown that similar arguments as above can also be
applied to non-unitary models where the spectrum may contain zero norm states
and therefore the scalar product used in conformal perturbation theory has to
be adapted to properly deal with Jordan cells, see e.g.\
Ref.~\onlinecite{DuJS10}.  With such a modification the perturbative
analytical corrections to scaling in a logarithmic minimal CFT with central
charge $c=-2$ describing critical dense polymers have been found to coincide
with the exactly known spectrum for a lattice model \cite{IzRH12}.

\begin{figure}[t]
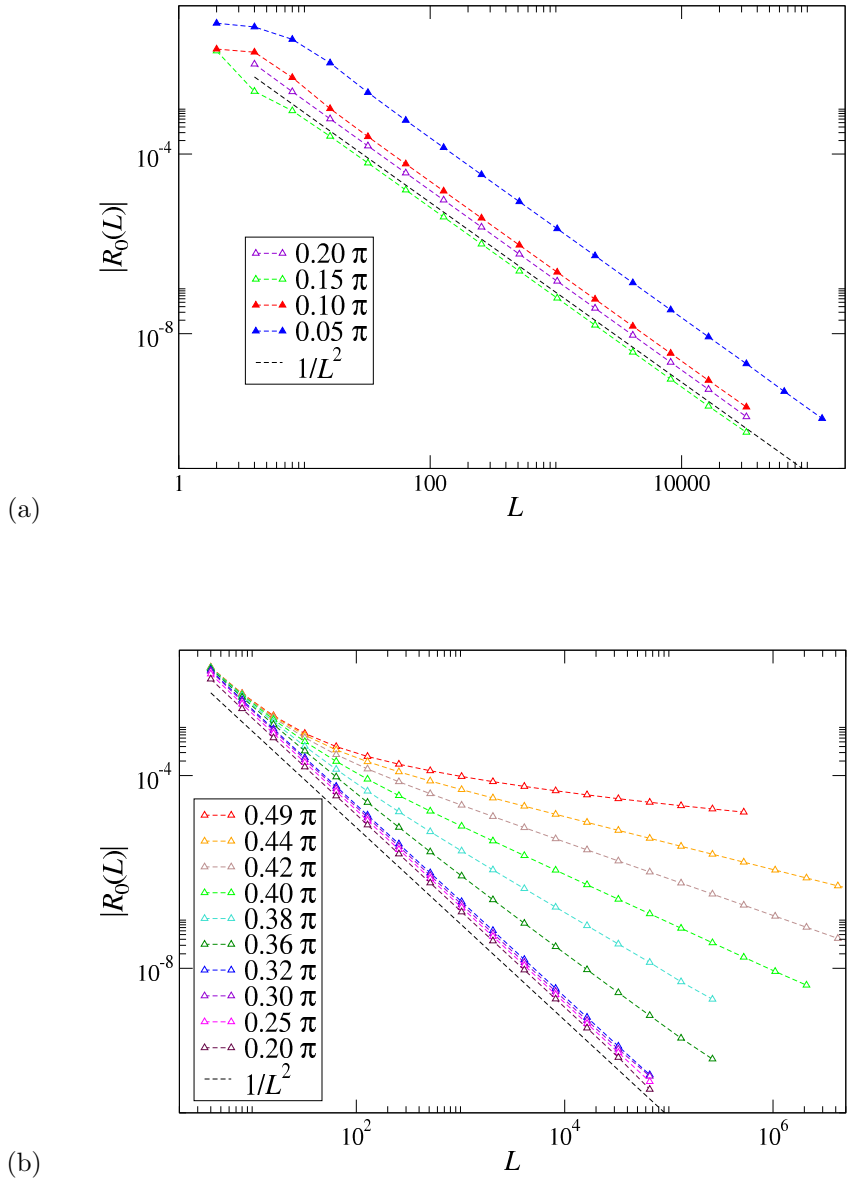

\begin{center}
(a) \hspace{5mm}
\includegraphics[width=0.6\textwidth]{corrsc3_p1q1_gs.eps}
\end{center}
\vspace*{2em}
\begin{center}
(b) \hspace{5mm}
\includegraphics[width=0.6\textwidth]{corrsc2_p1q1_gs.eps}
\end{center}
\caption{The scaling correction $R_0(L)$ to the ground state energy of the
  self-dual model for anisotropies \mbox{(a) $0<\gamma\leq0.2\pi$} and (b)
  $0.2\pi \leq \gamma < \frac\pi2$. Positive (negative) $R_0(L)$ are depicted
  by filled (open) symbols.}
\label{logfinestruct0}
\end{figure}

\begin{figure}[t]
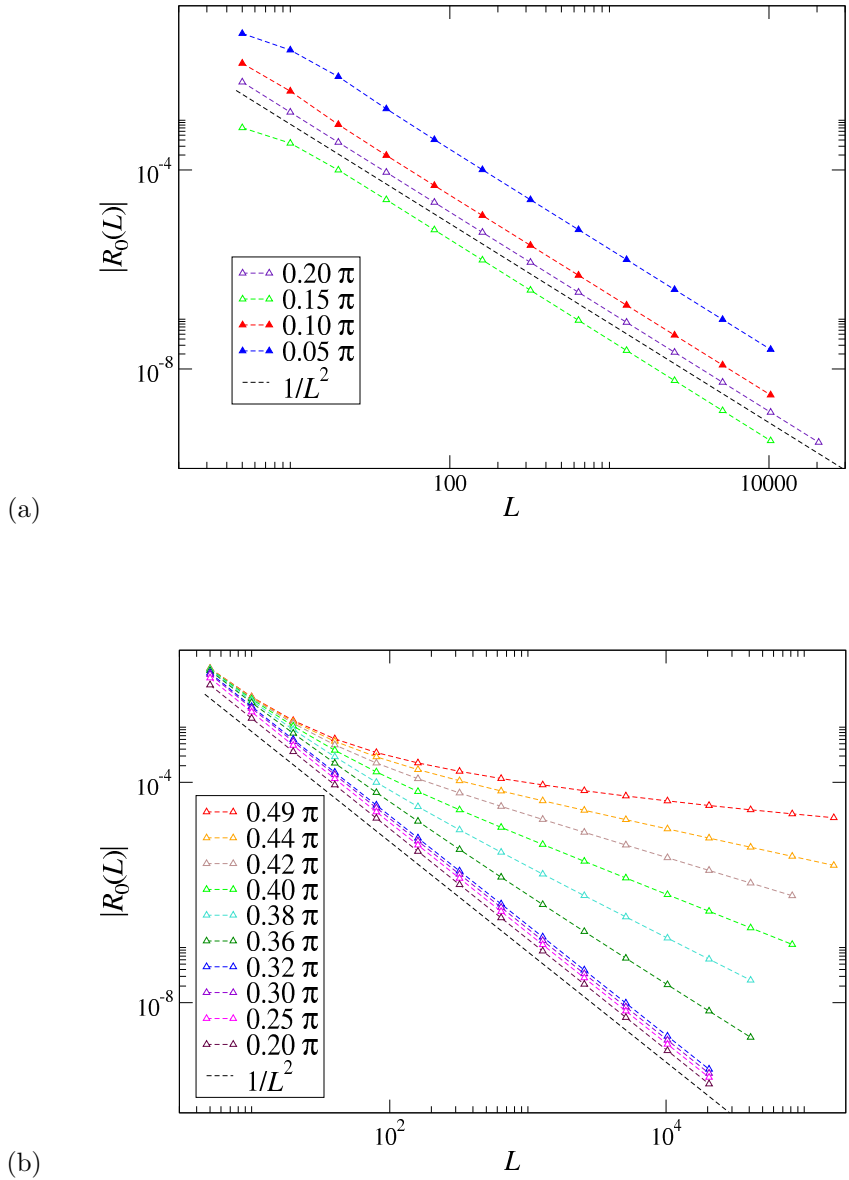

\begin{center}
(a) \hspace{5mm}
\includegraphics[width=0.6\textwidth]{corrsc1_p3q2_gs.eps}
\end{center}
\vspace*{2em}
\begin{center}
(b) \hspace{5mm}
\includegraphics[width=0.6\textwidth]{corrsc3_p3q2_gs.eps}
\end{center}
\caption{Same as Fig.~\ref{logfinestruct0} but for the model with staggering
  $\alpha(p=3,q=2)$.}
\label{logfinestruct1}
\end{figure}

\begin{figure}[t]
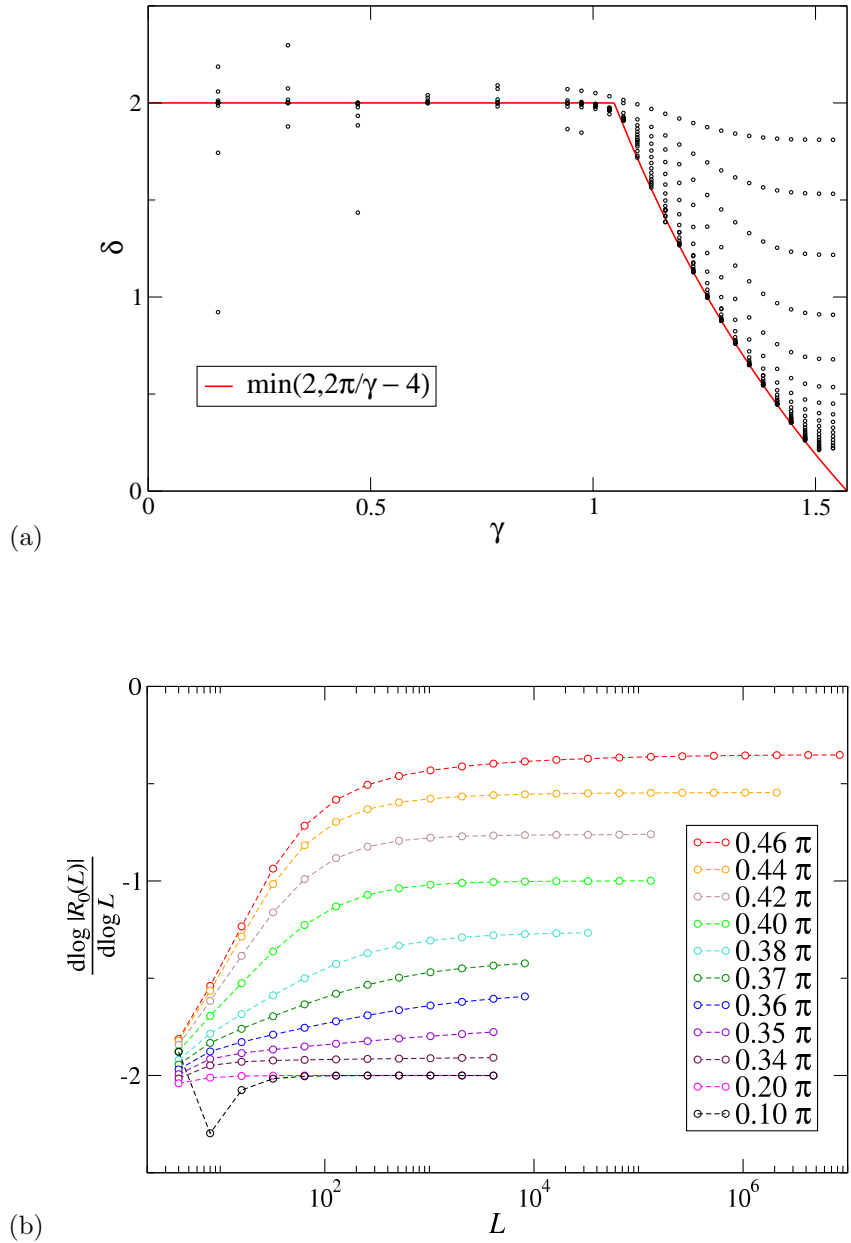

\begin{center}
(a) \hspace{5mm}
\includegraphics[width=0.6\textwidth]{exponents2_p1q1_gs.eps}
\end{center}
\vspace*{2em}
\begin{center}
(b) \hspace{5mm}
\hspace*{-1.6em}
\includegraphics[width=0.627\textwidth]{exponents3_p1q1_gs.eps}
\end{center}
\caption{(a) Conjectured $\gamma$-dependence (\ref{exponent}) of the exponents
  $\exponent$ in the algebraic decay of $R_0(L)$ (red line).  Symbols denote
  the finite size estimates obtained for the exponents from the ground state
  data of the self-dual model shown in Fig.~\ref{logfinestruct0}.  (b) System
  size dependence of the finite size estimates
  $\log\left(R_0(L)/R_0(L/2)\right)/\log2$ 
  converging to $-\exponent$ in the large-$L$ limit.  }
\label{exponents0}
\end{figure}

\begin{figure}[t]
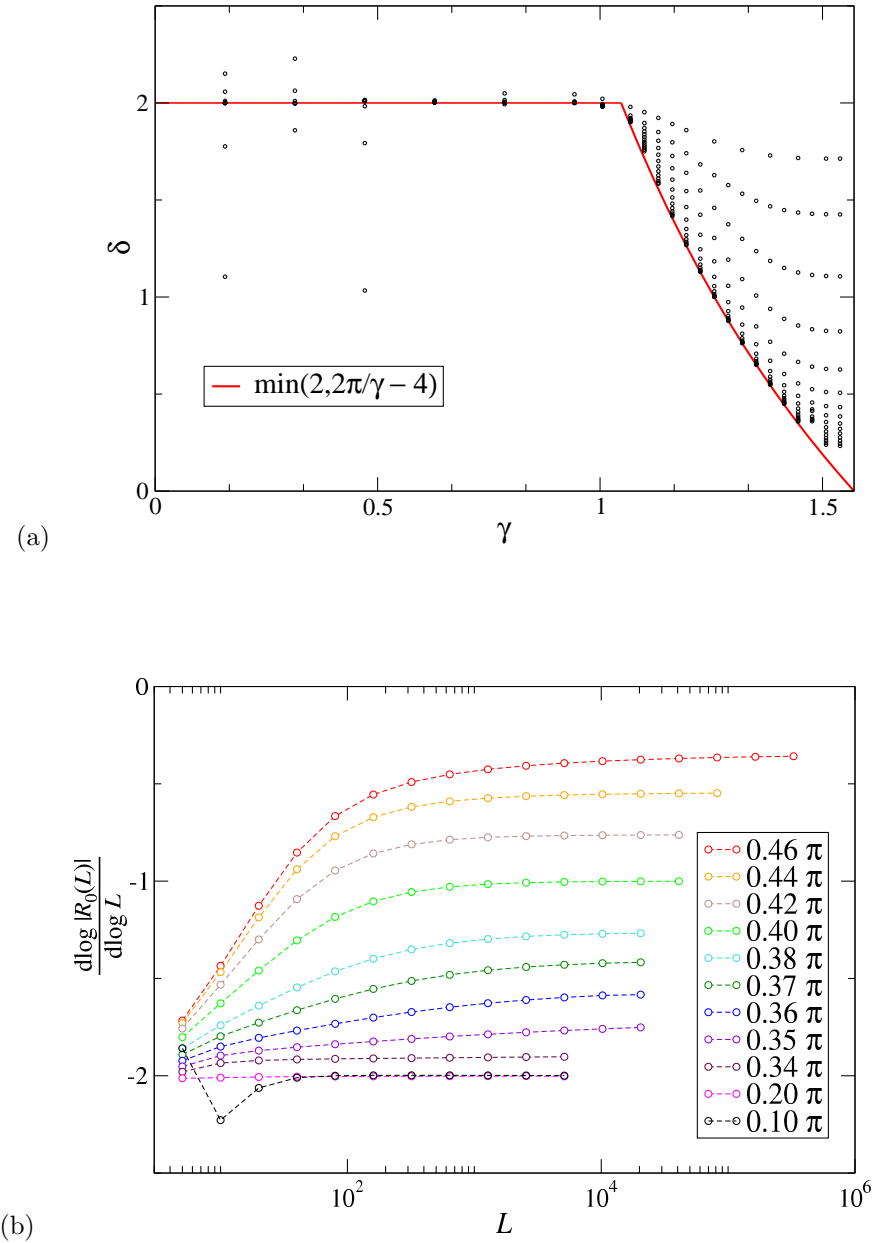

\begin{center}
(a) \hspace{5mm}
\includegraphics[width=0.6\textwidth]{exponents2_p3q2_gs.eps}
\end{center}
\vspace*{2em}
\begin{center}
(b) \hspace{5mm}
\hspace*{-1em}
\includegraphics[width=0.64\textwidth]{exponents3_p3q2_gs.eps}
\end{center}
\caption{Same as Fig.~\ref{exponents0} but refering to the data with staggering
  $\alpha(p=3,q=2)$ shown in Fig.~\ref{logfinestruct1}.}
\label{exponents}
\end{figure}

\begin{figure}[t]
\begin{center}
\includegraphics[width=0.7\textwidth]{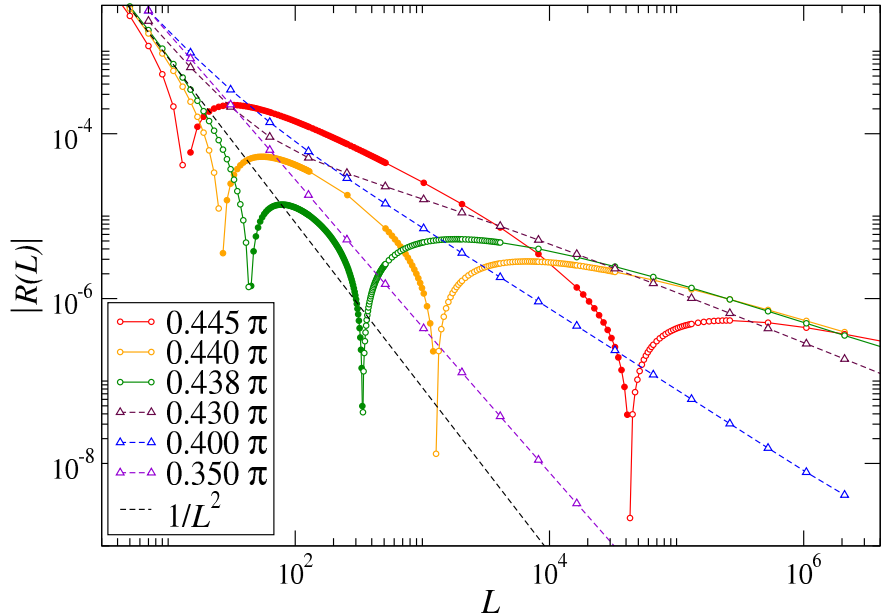}
\end{center}
\caption{Scaling correction $R(L)$ to the excited state ($(m,w)=(0,0)$ and
  $\widetilde{m}=1$, i.e.\ $n_1-n_2=1$) of the self dual model.  Positive
  (negative) $R(L)$ are depicted by filled (open) symbols.}
\label{oscillating00}
\end{figure}

%%%%%%%%%%%%%%%%%%%%%%%%%%%%%%%%%%%%%%%%%%%%%%%%%%%%%%%%%%%%%%%%%%%%%%
Corrections to scaling similar to (\ref{corrsc}) are expected to arise in
systems with a continuous spectrum \cite{Zamo06}.  In fact, our numerical
results for the corrections to the finite size scaling of the ground state of
the staggered six-vertex model, i.e.\ $(m,w)=(0,0)$ and $\widetilde{m}=0$,
indicate that $R_0(L)$ vanishes asymptotically as a power of the system size,
i.e.\ $R_0(L)\sim L^{-\exponent}$, see Figs.~\ref{logfinestruct0} and
\ref{logfinestruct1}.
%
%======================================================================%
%======================================================================%
%
%======================================================================%
%======================================================================%
%
For $\gamma\lesssim\pi/3$ the dominant term vanishes as $L^{-2}$ with an
amplitude which changes its sign for an anisotropy in
$0.1\pi\lesssim\gamma\lesssim 0.15\pi$.  For $\gamma>\pi/3$ we observe a slow
crossover from the $L^{-2}$ behaviour to a different power law which takes
place over several orders of magnitude in the system size. 
Based on numerical results for $L$ up to $10^6$ obtained by solving the NLIEs
we conjecture that the exponent $\exponent$ governing the asymptotic behaviour
of $R_0(L)$ (i.e.\ the threshold to the continuous spectrum above the lowest
state with $(m,w)=(0,0)$) is given by
\begin{equation}
  \label{exponent}
  \exponent=
  \begin{cases}
    \phantom{x}2 &\text{for} \quad 0 < \gamma < \frac\pi3\\
    \frac{2\pi}{\gamma}-4 &\text{for} \quad \frac\pi3\leq \gamma < \frac\pi2
  \end{cases}\,.
\end{equation}
This behaviour does not depend on the staggering $\alpha$, c.f.\
Figs.~\ref{exponents0} and \ref{exponents}.  
%
%======================================================================%
%======================================================================%
%
%======================================================================%
%======================================================================%
%
We also have analyzed the corrections to scaling in various excited states.
For excitations within the continuum above $(m,w)=(0,0)$ but $\tilde{m}\ne0$
we have used the NLIEs to compute the eigenvalues.  States with
$(m,w)\ne(0,0)$ are outside the range of validity of our NLIEs.  For these we
have solved the Bethe equations (\ref{BAequ}) directly which puts a limitation
on the available data to lattices with a few thousand sites.  In all cases
$R(L)$ changes its sign as a function of $L$ for certain values of the
anisotropy, see e.g.\ Fig.~\ref{oscillating00}.
%
%======================================================================%
%======================================================================%
%
Therefore, even larger system sizes are needed for a quantitative analysis of
the asymptotic behaviour of $R(L)$.  Based on our data, however, we find an
algebraic decay consistent with the conjecture (\ref{exponent}).

To interpret these findings for the staggered six-vertex model with its low
energy description in terms of the Euclidean black hole sigma model CFT
the considerations leading to (\ref{corrsc}) need to be modified as follows:
first of all, the presence of a continuous component in the spectrum of
critical exponents implies that the sums in (\ref{cftpert}) and (\ref{corrsc})
should be replaced by integrals and the coupling constants become functions
$g(s)$ of the continuous quantum number $s$.  We interpret the slow crossover
in the scaling behaviour of $R(L)$ from $L^{-2}$ to $L^{-\exponent}$ observed
for $\gamma>\pi/3$ as an indication for the presence of a perturbation of the
fixed point Hamiltonian by a continuum of conformal fields.
Furthermore, since the ground state of the lattice model is not the 
(non-normalizable) vacuum of the CFT but rather the state corresponding to the
lowest conformal weight $(h_0,\bar{h}_0)$, first order corrections cannot be
excluded to contribute to $R_0(L)$.  This is consistent with the fact that the
asymptotic behaviour of \emph{all} states considered is governed by the same
exponent (\ref{exponent}).

% first, as an immediate consequence of the continuous component in the spectrum
% of critical exponents the sums in (\ref{cftpert}) and (\ref{corrsc}) have to
% be replaced by integrals and the coupling constants become functions $g(s)$ of
% the continuous quantum number $s$ which may lead to large contributions from
% higher orders due to the presence of small denominators.

Taking into account these modifications to (\ref{corrsc}) would lead to the
conclusion that the numerical data for $R(L)$ for the lattice model are the
first order effect of a perturbation of the fixed point Hamiltonian by a
descendent of the identity operator with dimension $X_I=4$ and a continuum of
operators starting with dimension $X_k=2\pi/\gamma-2 = 2(k-1)$.  
The latter, however, is not in the spectrum (\ref{cftbh}) of the
$SL(2\mathbb{R})/U(1)$ coset model\footnote{This is also true if one considers
  the normalizable operators from the series of \emph{discrete}
  representations with real $SL(2,\mathbb{R})$ spin $j$ which can take values
  $1/2<-j<(k-1)/2$ subject to a constraint on physical states relating $j$ to
  the charge $m$ and vorticity $w$ \cite{MaOo01,HaPT02}.}:
while the $\gamma$-dependence of (\ref{exponent}) could be realized by a
perturbation through fields with vorticity $w=\pm2$ there is no sign of the
divergence due to the contribution of the non-compact degree of freedom to the
conformal weights as $\gamma\to\pi/2$ (or $k\to2$).

To resolve this discrepancy one has to consider additional ways how the
regularization of the CFT in terms of the staggered six-vertex model on a
finite lattice can affect the asymptotic $L$-dependence of the corrections to
scaling, Eq.~(\ref{corrsc}).  Here one has to take into account that the
latter are given -- apart from the coupling constants appearing in the
perturbation (\ref{cftpert}) -- in terms of universal quantities such as the
scaling dimensions and OPE coefficients of the CFT.

Let us now assume that the perturbation of the fixed point Hamiltonian $H^*$
present in the staggered six-vertex model is given in terms of an operator
from the conformal block of the identity with dimension $X_I=4$ and operators
from the continuum of fields with quantum numbers $(m,w)=(0,2)$ and
$SL(2,\mathbb{R})$-spin $j=(-1/2+\I s)$.  On a finite lattice the latter is
quantized as a consequence of (\ref{mtvss}) with $\Delta s\simeq \pi/({2}\log
L)$.  Then, for sufficiently large $L$, the OPE coefficients appearing to the
first order expression for the corrections to scaling in the ground state of
the lattice model $(m,w)=(0,0)$ and $s=0$ are
\begin{equation}
  \label{opebh}
  C_{(0,0),(0,0),(0,2)}\left(s_1=0,s_2=0,s_3=\frac{\pi n}{{2}\log L}\right)\,,
  \quad n=0,1,2,\ldots
\end{equation}
(here we have indicated the dependence of $C_{a,b,c}$ on the participating
fields with conformal weights (\ref{cftbh}) through subscripts for the
discrete quantum numbers $(m_a,w_a)$ while the continuous quantum numbers
$s_a$ appear as arguments of the OPE coefficient).  Little is known about the
operator product expansion in theories with a non-compact target space.  For
two systems related to the present model, i.e.\ Liouville field theory and the
$H_3^+$ Wess-Zumino-Novikov-Witten model, the OPE coefficients have been found
to be given in terms of double Gamma functions depending on combinations of
the spins $j_a=(-1/2+\I s_a)$ and $k=\pi/\gamma$ \cite{DoOt94,ZaZa96,Tesc00}.
With (\ref{opebh}) this gives a rise to an additional $L$-dependence in the
individual terms contributing to the corrections to scaling (\ref{corrsc})
which may account for the observed asymptotics with exponent (\ref{exponent}).

Finally, we note that the exponent (\ref{exponent}) vanishes as $\gamma$
approaches $\pi/2$ indicating the appearance of a marginal operator in the
perturbation of the fixed point Hamiltonian which leads to a different low
energy effective theory.  In the staggered six-vertex model some of the vertex
weights vanish in this limit and the lattice model has an $OSP(2|2)$ symmetry
\cite{IkJS08}.

%%%%%%%%%%%%%%%%%%%%%%%%%%%%%%%%%%%%%%%%%%%%%%%%%%%%%%%%%%%%%%%%%%%%%%
\section{Discussion}
We have investigated the finite size spectrum of the staggered six-vertex
model for the range of parameters $0\le\gamma<\alpha<\pi-\gamma$.
As has been noted in previous works the continuous component of this spectrum
leads to a strong logarithmic size dependence \cite{JaSa06,IkJS08,FrMa12}.
Therefore both a formulation of the spectral problem allowing for numerical
studies of large system sizes and insights into the parametrization of the low
energy degrees of freedom in terms of the parameters of the lattice model are
needed.  For the self-dual model, $\alpha=\pi/2$, these points have been
addressed before and provided evidence for the proposal that the critical
theory of the model is the $SL(2\mathbb{R})/U(1)$ sigma model at level
$k=\pi/\gamma>2$ describing a two-dimensional Euclidean black hole
\cite{IkJS12,CaIk13}.

We have derived a set of coupled NLIEs (\ref{aux1}) and (\ref{aux2}) which
generalize the ones obtained previously for the self-dual case $\alpha=\pi/2$
\cite{CaIk13} to the range of staggering given above.  The kernel functions
appearing in the NLIEs used here are regular in Fourier space.  As a
consequence this formulation is particularly suitable for their numerical
solution: we can compute the energies of the ground state and in the continuum
above it for chains with up to $10^6$ lattice sites for arbitrary staggering
$\gamma<\alpha\le\pi/2$.  Based on our numerical data we have extended the
proposal \cite{IkJS12} for the quantum number for the continuous part of the
spectrum in terms of the conserved quasi-momentum of the vertex model for
staggering away from the self-dual line, Eq.~(\ref{qnocont}).
With this input we were able to compute the density of states of the model
from the finite size spectrum obtained by numerical solution of the NLIEs.
Together with the existing data for the finite size spectrum \cite{FrMa12}
this shows that the model is in the same universality class as the self-dual
model independent of the staggering $\gamma<\alpha<\pi-\gamma$.  Both the
finite size spectrum and the density of states agree with what is known about
the Euclidean black hole sigma model.

Finally, we have extended previous studies of the finite size spectrum
\cite{IkJS08,FrMa12} by considering the corrections to scaling due to
irrelevant perturbations of the fixed point Hamiltonian appearing in the
lattice model.  Such perturbations are expected to lead to subleading
power-laws in the finite size spectrum which can provide additional
information on the operator content of the continuum model and insights into
the emergence of the continuum of critical exponents in the thermodynamic
limit of the lattice model.  Again, our numerical data suggest that the
variation of the staggering parameter does not change the critical theory:
different values $\alpha$ only lead to small changes in the non-universal
coupling constants $g$ in (\ref{cftpert}).  As for the interpretation of our
numerical results summarized in the conjecture (\ref{exponent}) for the
asymptotic algebraic decay of the corrections to scaling, however, we find
that the known predictions for theories with purely discrete spectrum have to
be modified here.  These modifications appear to be closely related to the way
how the non-compact degree of freedom is dealt with in the regularization of
the field theory leading to the staggered six-vertex model.  To make progress
in this direction additional work from the CFT side is called for, in
particular with respect to the operator product expansion in theories with
non-compact target space.

A natural extension to our work would be the finite size scaling analysis of
the lattice model in sectors with non-zero magnetization, i.e.\ with
$m=n_1+n_2-L\ne 0$, or non-zero vorticity $w$.  In the derivation of the
corresponding NLIEs this amounts to consider hole-type solutions of the Bethe
equations (\ref{BAequ}) appearing inside the integration contours which lead
to additional logarithmic driving terms.
Another direction for future work is to consider more general lattice models
which develop a continuous spectrum of critical exponents in the thermodynamic
limit.  Known examples with such a behaviour are the supersymmetric vertex
models based on alternating representations of $U_q[gl(2|1)]$.  These models
are known to contain the staggered six-vertex model studied in the present
work as a subsector \cite{EsFS05,FrMa11,FrMa12}.  Apart from general insights
into the critical properties of quantum spin chains based on super Lie
algebras and conformal field theories with non-compact target space this may
also provide a basis for an improved understanding of some topical problems in
condensed matter physics, e.g.\ the quantum phase transitions in
two-dimensional disordered systems \cite{MaNR98,Gruz99,Gade99,Zirn99} or
possibly the deconfinement of $U(1)$ gauge fields coupled to the Fermi
surface of a two-dimensional system \cite{Lee09,MeSa10}, in the context of an
exactly solvable model.

\begin{acknowledgments}
  We thank M\'arcio J. Martins and Jan Hinnerk Stosch for useful discussions.
  This work has been supported in part by the Deutsche Forschungsgemeinschaft.
\end{acknowledgments}
%%%%%%%%%%%%%%%%%%%%%%%%%%%%%%%%%%%%%%%%%%%%%%%%%%%%%%%%%%%%%%%%%%%%%%%

% \bibliographystyle{amsplain}
% \bibliography{base,bound,frahm,books,s6v}
\providecommand{\bysame}{\leavevmode\hbox to3em{\hrulefill}\thinspace}
\providecommand{\MR}{\relax\ifhmode\unskip\space\fi MR }
\providecommand{\MRhref}[2]{%
  \href{http://www.ams.org/mathscinet-getitem?mr=#1}{#2}
}
\providecommand{\href}[2]{#2}

\end{document}